\documentclass[usenatbib,onecolumn]{mn2e}
\usepackage{amsmath}
\usepackage{amsfonts}
\usepackage{amssymb}
\usepackage{natbib}
\usepackage{epsfig}

\title[Constraints on Radial Migration in Spiral Galaxies. I]{Constraints on Radial Migration in Spiral Galaxies I. Analytic Criterion for Capture at Corotation}
\author[Kathryne J. Daniel and Rosemary F. G. Wyse]{Kathryne J. Daniel$^{1}$\thanks{E-mail:kdaniel@jhu.edu}\thanks{Also known as Kathryne J. D. Tolfree} and Rosemary F. G. Wyse$^{1}$\\
$^{1}$Department of Physics \& Astronomy, Johns Hopkins University, Baltimore, MD 21218, USA}

\begin{document}
\date{Accepted 16 December 2014. Received 11 December 2014; in original form 28 October 2014}

\pagerange{\pageref{firstpage}--\pageref{lastpage}} \pubyear{2014}

\maketitle

\label{firstpage}

\begin{abstract}
Near the corotation resonance of a transient spiral arm, stellar orbital angular momenta may be changed without inducing significant kinematic heating, resulting in what has come to be known as radial migration. When radial migration is very efficient, a large fraction of disk stars experiences significant, permanent changes to their individual orbital angular momenta over the lifetime of the disk, having strong implications for the evolution of disk galaxies.  The first step for a star in a spiral disk to migrate radially is to be captured in a \lq\lq trapped" orbit, associated with the corotation resonance of the spiral pattern.  An analytic criterion for determining whether or not a star is in a trapped orbit has previously been derived only for stars with zero random orbital energy in the presence of a spiral with fixed properties.  In this first paper in a series, we derive an analytic criterion appropriate for a star that is on an orbit of finite random orbital energy.  Our new criterion demonstrates that whether or not a star is in a \lq\lq trapped'' orbit primarily depends on the star's orbital angular momentum. This criterion could be a powerful tool in the interpretation of the results of N-body simulations.  In future papers of this series, we apply our criterion to explore the physical parameters important to determining the efficiency of radial migration and its potential importance to disk evolution.
 
\end{abstract}

\section{Introduction}

Galaxies are known not to evolve in isolation, but rather under the influence of external effects such as interactions with neighbours, or gas infall/outflow. In addition, galaxies are also subject to secular internal processes \citep[see][and references therein]{Sellwood13,Kormendy13}.   One such process, resulting in what is now commonly referred to as \lq\lq radial migration", was identified by \cite{SB02}.  They demonstrated, using both analytic and numerical techniques, that under certain circumstances a transient spiral arm in a two-dimensional disk can permanently change the orbital angular momentum of a star  without causing a significant change in the star's orbital random energy. \cite{SB02} showed that in the epicyclic approximation, the increase in orbital random energy of a star, $\Delta E_{ran}$, due to an encounter with a spiral of pattern speed $\Omega_p$,  is related to its change in orbital angular momentum, $\Delta L_z$, as follows:
\begin{equation}\label{eqn:SB1}
\Delta E_{ran} \propto(\Omega_g-\Omega_p)  \Delta L_z
\end{equation}
where $\Omega_g$ is the orbital frequency of the star's guiding centre. Thus a star that is on a nearly circular orbit with guiding centre radius approximately equal to the corotation radius ($R_{CR}$, at which the circular orbital frequency, $\Omega_c(R_{CR})$, equals the pattern speed of the spiral) can experience a finite change in orbital angular momentum with negligible accompanying increase in random motions.

The change in orbital angular momentum corresponds to a change in guiding centre radius. As proposed by \cite{SB02}, such radial excursions induced through scattering across corotation by a \textit{transient} spiral could be permanent, leading to \lq\lq radial migration", whereby stars move radially within the disk without experiencing a significant change in their orbital circularity, i.e.~the stellar disk remains kinematically cold.  Note this restricted application of the term \lq\lq radial migration'', excludes a change of guiding centre radius that is associated with heating; there is no consensus on the usage of this term in the literature, but was named \lq\lq churning" by \cite{SB09a}. 

The importance of radial migration to galaxy evolution depends on both the duty cycle for transient spiral arms and the efficiency of radial migration from each transient spiral arm.  Should a disk's history include the occurrence of multiple, transient spiral arms with a distribution of pattern speeds (and hence a distribution of corotation radii across the extent of the disk), radial mixing of stellar populations - and of gas - due to the induced radial migration could have a substantial impact on the chemical, kinematic and structural evolution of disk galaxies \citep[e.g.][]{SB02,Debattista06,Sellwood13}.  However, the physical parameters that determine the efficiency of radial migration, and therefore its importance to galaxy evolution, have not yet been rigorously explored.

\cite{SB09a,SB09b} pioneered an exploration of how radial migration could affect disk evolution by including a prescription for the probability of radial migration within a chemical evolution model for a spiral galaxy.  In their model they assumed disks of gas and stars with fixed exponential scale lengths and derived an expression for the probability for radial migration of stars and gas for a given annulus of the disk that scaled simply with the surface mass density of that annulus.  This probability was independent of the properties of the spiral perturbations or of the stellar populations.\footnote{Their expression for the probability of radial migration was based on a dimensional argument considering a random walk within a  self-gravitating disk.}  \citeauthor{SB09a} assumed that the vertical energy of a migrating stellar population (characterised by the vertical velocity dispersion, $\sigma_z$) is conserved.  Their model led to the emergence of a stellar thick disk-like structure as stars that migrated outward from the inner regions of the disk, where velocity dispersions are higher, experienced a weaker vertical restoring force from the lower surface-density outer disk.  However, in a 3D N-body simulation, \cite*{SSS12} found that it is not the vertical energy of a population that is conserved but rather the vertical action and that outwardly migrating stars do not thicken the disk enough to produce a thick disk-like structure \citep[see also][]{Minchev12b}.

One might appeal to high-resolution N-body/SPH simulations for clarification on how radial migration could affect disk evolution.  However, such simulations published to date show evidence for varying degrees of importance for radial migration, ranging from negligible \citep*{BKW12} to significant, with radial migration arguably being responsible for populating the outer disk with old stars \citep{Roskar08a} and for the formation of a thick disk-like structure \citep{Loebman11}.  The origin of this range in the inferred importance for radial migration to disk evolution is poorly understood, in large part because of the many parameters in the simulations and the fact that it is unknown what physical parameters underlie the efficiency of radial migration.  Thus, it is uncertain how time- and mass-resolution, integration time, the presence or not of a live, dark-matter halo, the initial conditions and assumed sub-grid physics may affect the evolution of simulated disks in this respect \citep[see also][]{Sellwood13}.  As a result, it is unclear how the efficiency of radial migration in these simulations relates (or not) to that of real disk galaxies.

In principle, there should be observable signatures of the past efficiency of radial migration in the properties of solar neighbourhood stars, such as an age-dependent spread in the local age-metallicity relationship \citep{FM93,SB02}. In practice, however, observational uncertainties currently make it difficult to determine the amplitude (if any) of such signatures.

Analytical arguments offer an alternate means to gain insight into the physical parameters important to the efficiency of radial migration.  The published analytic derivations of the requirements for radial migration offer limited insight because they are applicable only to stars on purely or nearly circular orbits \citep{Contopoulos78,BT87,SB02}. It is generally expected that the effectiveness of radial migration must decrease with increasingly non-circular orbits, since stars in non-circular orbits cannot keep station with the perturbing pattern speed \citep{SB02}.  Indeed, \cite{SSS12} found a decrease in the RMS change in the orbital angular momenta of a population of star particles in the presence of a transient spiral, as the value of the radial velocity dispersion ($\sigma_R$) of the population increased (see their Fig.~10).  In a separate N-body simulation of a sub-maximal disk, \cite{Vera-Ciro14} found that radial migration preferentially affected stellar populations with low vertical velocity dispersion.  The impact of spiral strength, lifetime, and pitch angle on the efficiency of radial migration remains largely unexplored (but will be addressed in this study).

This series of papers is intended to identify the important factors contributing to the efficiency of radial migration and to determine simple scaling relations which can be implemented within models of disk evolution.  Our analysis of the efficiency of stellar radial migration is simplified to only include migration that is induced by a single spiral pattern in a 2D disk.  It should be noted that gas is also affected at the corotation resonance, but we do not address the migration of gas or, indeed, any dissipative physics here.  We use the following three measures of the efficiency of radial migration: (1) the fraction and distribution of disk stars captured in trapped orbits; (2) the RMS change in orbital angular momentum without associated heating of migrating stars; and (3) the time-scale to reach the  maximum value of the RMS change in orbital angular momentum for the ensemble of migrating stars.  In this first paper, we derive an analytic criterion to determine whether or not a star with some finite random orbital energy is captured in a trapped orbit.  A second paper will focus on measure (1) by applying this criterion to models of stellar populations with a given distribution function in a disk galaxy.  A third paper in this series will discuss measures (2) and (3).  The final paper will use the scaling relations from papers~I,~II and~III to address how radial migration could affect the evolution of spiral galaxies with different assumed histories of spiral perturbations.

This paper is organized as follows.  In \S\ref{sec:Capture}, we first review in detail the existing analytic description of the requirements for whether or not a disk star is in a trapped orbit (also defined in \S\ref{sec:Capture}) and then we derive a new analytic criterion that is applicable to stars that are on non-circular orbits.  In \S\ref{sec:NumericalExploration} we use an orbital integrator to test - and confirm - the robustness of this new criterion. In \S\ref{sec:Discussion}, we briefly discuss the consequences of a non-steady spiral pattern.  Section~\ref{sec:Conclusions} presents a summary of our conclusions.

\section{Requirements for Capture into a Trapped Orbit}\label{sec:Capture}

The first step of the process that can lead to radial migration is for an object (star, asteroid, planet, etc.) to be captured onto a particular family of resonant orbits that can occur near the radius of corotation with a non-axisymmetric potential.  Members of this family of orbits are uniquely characterized by periodic changes in orbital angular momentum with negligible accompanying change in random orbital energy.  A star on an orbit that is a member of this family will henceforth be referred to as being in a \lq\lq trapped" orbit \citep[the name proposed by][]{Barbanis76}.  Trapped orbits were predicted as a solution to the three-body problem for an asteroid in the orbital path of Jupiter around the Sun \citep{Brown11}, and were dubbed \lq\lq tadpole" or \lq\lq horseshoe" orbits, because these approximate the shape of the orbital path in the rotating frame \citep[see][for a discussion of these orbits in planetary systems]{GT82}.  The shapes of stellar orbital paths about the galactic centre in the presence of a spiral or bar perturbation are slightly different, but the physics is the same (see below).

\subsection{Underlying Physics}\label{sec:Underlying}

In order to understand the periodic changes in orbital angular momentum of a star in a trapped orbit, it is instructive to restrict the analysis to an initially circular orbit in a 2D disk with a flat rotation curve.  In this case the star's initial orbital radius, $R$, equals its guiding centre radius, $R_g$, and its azimuthal velocity, $v_\phi$, equals the circular velocity, $v_c$. Assume that a steady $m$-armed perturbation to the potential is imposed, with pattern speed $\Omega_p=v_c/R_{CR}$ and that the star is located close to  the radius of corotation, $R_{CR}$. Further, assume that the strength of the perturbation is approximately constant across corotation.  The net force on the star, in the frame that rotates with the pattern is then predominantly in the azimuthal direction ($\mathbf{F} \approx F_\phi \boldsymbol{\hat{\phi}}$).  The resulting torque ($\boldsymbol{\tau}=d \mathbf{L_z}/dt \propto R_{CR} F_\phi \mathbf{\hat{z}}$) will alter the star's $z$-directional angular momentum ($L_z$), which results in a change in the star's orbital radius.  The time-dependent torque a star experiences during a trapped orbit \citep[see the physical explanation in][ \S3.4]{SB02} will cause the star's guiding center to oscillate radially.  Should the perturbation be transient, the star will likely not complete this oscillation, its orbital angular momentum and guiding centre radius will then be permanently changed \citep{SB02} and the star will have migrated radially.

The analysis of orbits in a steadily rotating, non-axisymmetric disk is most conveniently carried out in the frame rotating with the pattern. As discussed in standard textbooks e.g.~\citet{BT08}, in such a potential neither energy nor angular momentum (measured in the non-rotating frame) is conserved, and consequently, there are no circular orbits in a non-axisymmetric potential.  The combination of orbital energy  in the inertial frame, $E$, and orbital angular momentum ($L_z$) given by the Jacobi integral, $E_J$, is conserved, where \citep[][eqns.~3.113]{BT08} 
\begin{equation}\label{eqn:EJ}
E_J = E - \Omega_pL_z = \frac{1}{2}|\dot{\mathbf{x}}|^2 + \Phi_{eff}(\mathbf{x}) ,
\end{equation}
with
\begin{equation}\label{eqn:EffectivePotential}
\Phi_{eff}(\mathbf{x}) \equiv \Phi(\mathbf{x}) -\frac{1}{2}|\mathbf{\Omega_p} \times \mathbf{x}|^2
\end{equation}
being the effective potential \citep[][eqns.~3.114]{BT08}, $\Phi(\mathbf{x})$ the actual potential, and $\mathbf{x}$ and $\mathbf{\dot{x}}$ being, respectively, the position and velocity of the star in the rotating frame. It is evident that a star with a given value of $E_J$ will have zero velocity at locations where $\Phi_{eff}(R,\phi) = E_J$, and thus the contour of $\Phi_{eff}$ with this value traces the Zero Velocity Curve (ZVC) for such stars. 

\begin{figure}
\begin{center}
\includegraphics[scale=1]{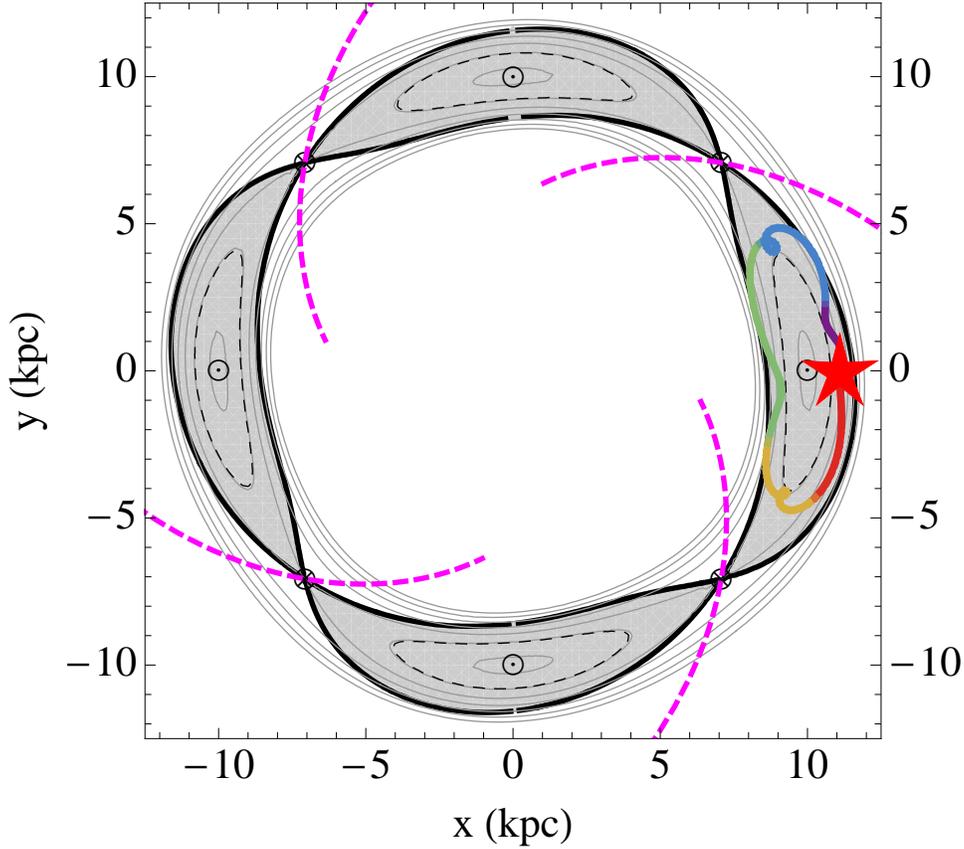}
\caption{Effective potential, $\Phi_{eff}$, for a trailing, $m~=~4$ spiral pattern with pitch angle $\theta~=~35^\circ$, $R_{CR}=10$~kpc and amplitude at corotation $|\Phi_s(R_{CR})|=322$~km$^2$~s$^{-2}$ superposed on an underlying logarithmic potential with $v_c=220$~km~s$^{-1}$.  The peak of the spiral perturbation is shown as thick, dashed magenta curves.  The local maxima in $\Phi_{eff}$ (between spiral arms) are marked with the symbol $\odot$ and the saddle points (the deepest part of the spiral potential at corotation) are marked with $\otimes$.   The capture region has a thick, black outline and is shaded grey.  The orbital path in the rotating frame of a captured star (shown in solid rainbow colours that begin red and end violet) that was launched with $(v_R,v_\phi)=(0,220)$~km~s$^{-1}$ in the inertial frame with initial position 1.1~kpc outside corotation and $\phi=0$ (marked with a red star) was followed for $1$~Gyr.  The ZVC for the star, where $\Phi_{eff}(R,\phi)=E_J$, are indicated by the dashed line. }
\label{fig:CR8_PhiEff}
\end{center}
\end{figure}

Figure~\ref{fig:CR8_PhiEff} shows an example of contours of $\Phi_{eff}(\mathbf{x})$ near the radius of corotation of a $m=4$~armed spiral pattern superposed on an underlying logarithmic potential (we describe our model in detail in \S\ref{sec:NumericalExplorationModel}).  The parameters are chosen such that for the spiral $R_{CR}=10$~kpc, pitch angle $\theta=35^\circ$ (measured from the line of azimuth, such that spirals with small $\theta$ are ring-like), amplitude at corotation $|\Phi_s(R_{CR})|=322$~km$^2$~s$^{-2}$, and with circular speed in the underlying potential $v_c=220$~km~s$^{-1}$.  The zero of the azimuthal coordinate, $\phi$, is taken to be halfway between the maxima of the spiral pattern and in the figure is along the direction of the positive $x$-axis, and $\phi$ increases in a counter-clockwise direction.  There exist $m=4$ local maxima in the effective potential at $R=R_{CR}$ and with azimuthal coordinates located at the mid-points between the spiral arms, i.e. at $\phi=\{0,\pi /2, \pi , 3 \pi /2\}$.\footnote{These local maxima correspond to the Lagrange points $L_4$ and $L_5$ in the three-body problem.}  There also exist $m=4$ local minima (saddle points) in the effective potential at $R_{CR}$ and $\phi= \{\pi /4, 3\pi /4, 5\pi /4, 7\pi /4\}$, at the peak of the spiral potential.  We define  $\phi_{max}$ and $\phi_{min}$ to be the azimuthal coordinates of the maximum and minimum of the effective potential nearest the star to be considered, respectively.

As may be seen in Fig.~\ref{fig:CR8_PhiEff}, the local maxima in the effective potential are surrounded by contours in the effective potential that encircle these maxima, rather than encircling the galactic centre.  The solid, black boundary of the grey shaded region in Fig.~\ref{fig:CR8_PhiEff} is where the Jacobi integral of a star with zero random energy in the inertial frame will be equal to the value of the effective potential at the local minima at the radius of corotation, and very nearly follows the closed contours of the effective potential that encircle the local maxima.  The shaded region will henceforth be called the \lq\lq capture region" for reasons that will become apparent below. The areas enclosed by the contours around the maxima in the effective potential and of the capture region are larger for higher amplitude spiral patterns (we explore how this affects the efficiency of radial migration in Paper~II); note that we have chosen a high-amplitude spiral for the purpose of illustration.

\cite{SB02} (their eqns.~1-4) used the conservation of $E_J$ (eqn.~\ref{eqn:EJ} for a star moving in a 2D non-axisymmetric potential to derive the relationship, in the epicyclic approximation, between the change in a star's orbital random energy, expressed in terms of the radial action, $J_R$, and the change in its orbital angular momentum.
They found that
\begin{equation}\label{eqn:SB02JR}
\Delta J_R = \dfrac{\Delta E_{ran}}{\kappa} = \dfrac{(\Omega_p-\Omega_g)}{\kappa} \Delta L_z ,
\end{equation}
where $\kappa$ is the epicyclic frequency of radial oscillations for a star with small excursions from a circular orbit \citep[][eqn.~3.80]{BT08},
\begin{equation}\label{eqn:kappa}
\kappa^2(R_L)=\left( R\dfrac{d\Omega_c^2}{dR}+4\Omega_c^2\right)_{R_L} ,
\end{equation}
and $R_L$ is the guiding centre radius.  The random orbital energy in the inertial frame, $E_{ran}=E_*-E_c(R_L)$, is evaluated in the underlying axisymmetric potential, where $E_*$ is the total stellar energy in the inertial frame, and $E_c(R_L)$ is the energy of a star in a circular orbit at $R_L$ \citep{Dehnen99}.  Clearly, as discussed by \cite{SB02}, when $\Omega_p = \Omega_g$ changes in $L_z$ are not accompanied by changes in random energy.

The orbital path of a star with nearly zero random orbital energy that is captured in a trapped orbit is visualized in Figure~\ref{fig:CR8_PhiEff}.  The star has initial position 1.1~kpc beyond corotation (causing it to lag the pattern in the rotating frame), mid-way between trailing spiral arms, which places it inside the capture region at $t=0$.  The trajectory of the star is plotted in rainbow colours, beginning with red and ending with violet.  The star has a low value of the initial velocity in the rotating frame and since $\Omega_g\approx \Omega_p$, little increase in random motions (see Eqn.~\ref{eqn:SB02JR}). The star therefore remains on a trajectory that closely follows its ZVC, the contour for $\Phi_{eff}(\mathbf{x})=E_J$ (dashed contour in fig.~\ref{fig:CR8_PhiEff}).  Note that the star sensibly does not cross into the forbidden region where $\Phi_{eff}(\mathbf{x})>E_J$.

Any star on an orbit with Jacobi integral less than the value of the effective potential at the local saddle point (the lowest value of the effective potential within the shaded grey area in Fig.~\ref{fig:CR8_PhiEff}) i.e.~$E_J < \Phi_{eff}(R_{CR},\phi_{min})$, will not be able to cross the corotation radius and will therefore not be able to oscillate around the local maximum in $\Phi_{eff}$, i.e. is not trapped and will instead have an orbital path in the rotating frame that encircles the galactic centre.

In contrast, a star with $E_J>\Phi_{eff}(R_{CR},\phi_{min})$ can cross the corotation radius.  Should such a star have zero random orbital energy in the inertial frame, i.e. the orbital energy of a star in circular orbit in the underlying axisymmetric potential at that radius, it would have coordinates that are inside the capture region (shown in the next section).  Its orbital path would closely follow a contour in $\Phi_{eff}(\mathbf{x})$ that encloses a local maximum, as shown in Fig.~\ref{fig:CR8_PhiEff}.\footnote{In the special case that a star with zero random energy has $E_J=\Phi_{eff}(R_{CR},\phi_{min})$, it can cross $R_{CR}$ and may be in a heteroclinic orbit \citep[e.g][]{Martinet74}.}

\subsection{Capture Criterion: Stars with Zero Orbital Random Energy}\label{sec:CaptureWithout}

The criterion for whether or not a star \textit{with zero random energy} in the inertial frame is captured in a trapped orbit (henceforth called the \lq\lq capture criterion") has been previously explored in the literature \citep{Contopoulos73,Contopoulos78,Papayannopoulos79a,Papayannopoulos79b,SB02,BT08}.  For both the reader's convenience and to highlight the physics relevant to this study, we will briefly summarize these analyses, but defer to those papers for a more thorough treatment.  The requirements for a star to be in a trapped (or \lq\lq horseshoe") orbit that are derived in \cite{BT87} (Ch.~3.3b) and discussed in \cite{SB02} use a different set of assumptions from the capture criterion outlined in this section.  In appendix~\ref{sec:AppxA}, we explore the relationship between these two criteria.\footnote{The capture criterion of \cite{BT87} emerges from the equations of motion for a star in a trapped orbit. \cite{SB02} used the equations of motion from this analysis to approximate the maximum radial excursion during, and the minimum peroid for, a trapped orbit (their equations (11) and (12)).}

\cite{Contopoulos78} derived an analytic capture criterion for stars with zero random energy in a 2D disk where the potential, $\Phi(R,\phi)$, was composed of an underlying axisymmetric potential, $\Phi_0(R)$, plus a perturbation, $\Phi_1(R,\phi)$, that varied sinusoidally in the azimuthal direction.  In the frame rotating with the pattern, the perturbation to the potential is given by:
\begin{equation}
\Phi_{1}(R,\phi) = |\Phi_{s}(R)| \cos (m\phi),
\end{equation}
with $|\Phi_{s}(R)|$ being the amplitude of the potential and $m$ the number of spiral arms.

\citeauthor{Contopoulos78} simplified the equations by introducing the constant quantity  denoted here by $h_{CR}$, being  the value of the Jacobi integral for a star in a circular orbit at $R_{CR}$ in the underlying axisymmetric potential.  He then used a $2^{nd}$ order expansion of the Jacobi integral ($E_J$) of a star near the maximum in $\Phi_{eff}$, at $(R_{CR},\phi_{max})$, in terms of action-angle variables \citep[re-expressed by][eqn.~37]{Papayannopoulos79b},

\begin{equation}\label{eqn:EJexpanded}
E_J=h_{CR}+\kappa_{CR} J_R+a_{CR} J_R^2+2b_{CR} J_R J_\phi+c_{CR} J_\phi^2 +\Phi_{s,CR}\cos(m \phi_1) ,
\end{equation}
where the coefficients \citep[][Appx.~A]{Contopoulos75},
\begin{equation}\label{eqn:coeffs}
\begin{array}{cl}
a_{CR}&=\dfrac{1}{16} \kappa_{CR}^2 \left[\Phi_{0,CR}''''+\dfrac{60 \Phi_{0,CR}'}{R_{CR}^3}-\dfrac{5}{3\kappa_{CR}^2}\left(\Phi_{0,CR}'''-\dfrac{12\Phi_{0,CR}'}{R_{CR}^2}\right)^2\right]\\
b_{CR}&=\dfrac{\Omega_{CR} \kappa_{CR}'}{R_{CR}\kappa_{CR}^2}\\
c_{CR}&=\dfrac{\Omega_{CR}\Omega_{CR}'}{R_{CR} \kappa_{CR}^2}
\end{array}
\end{equation}
where prime indicates the radial derivative, $\phi_1$ is the azimuthal angular distance from $(R_{CR},\phi_{max})$ (i.e. $\phi_1=\phi-\phi_{max}$), $J_\phi=L_z(R)-v_{circ}(R_{CR})R_{CR}$ is the azimuthal action in the rotating frame and the subscript \lq\lq$CR$" denotes evaluation at the radius of corotation.

Setting $J_R=0$ in eqn.~\ref{eqn:EJexpanded}, as
appropriate for a star with zero orbital random energy, yields a simple quadratic equation:
\begin{equation}
J_\phi^2=\frac{1}{c_{CR}}\left[ E_J-h_{CR}-|\Phi_s|_{CR}\cos(m\phi_1) \right] .
\end{equation}
As \citeauthor{Contopoulos78} noted, real solutions (requiring that the right-hand side be greater than zero) exist for all values of the angle $\phi_1$ only if $E_J-h_{CR} <- |\Phi_s|_{CR}$ \citep[eqn.~16]{Contopoulos78}.  Real solutions exist for only a restricted range in $\phi_1$ when \citep[see also][eqn.~17]{Contopoulos78}
\begin{equation}\label{eqn:Contopoulos78}
-|\Phi_s|_{CR} < E_J-h_{CR} \leq |\Phi_s|_{CR}  .
\end{equation} 
Such stars are captured in trapped orbits, oscillating about the maximum in the effective potential (recall that the approximate expression for the Jacobi integral was valid only near the maximum).

Both $E_J$ and $h_{CR}$ are time-independent, so that the quantity,
\begin{equation}\label{eqn:Lambda_c}
\Lambda_c \equiv \dfrac{E_J-h_{CR}}{|\Phi_s|_{CR} } 
\end{equation}
is also conserved, (where the subscript \lq\lq $c$" denotes that the analysis assumes a star on a circular orbit in the underlying axisymmetric potential). Substitution for this parameter reduces the statement of the criterion for a star with zero random energy to be trapped (eqn.~\ref{eqn:Contopoulos78}) to the more compact form:
\begin{equation}\label{eqn:trapped}
-1< \Lambda_c \leq 1 .
\end{equation}

A star (with zero orbital random energy) that meets the capture criterion - and is therefore in a trapped orbit - could migrate radially if the perturbation were transient.  Should the star have random energy, however, eqn.~\ref{eqn:trapped} would no longer be a valid criterion to determine whether or not the star is in a trapped orbit.  We turn to this more realistic situation in \S\ref{sec:DerivationWRanMotions}.

\subsubsection{Capture Region}\label{sec:CaptureRegion}

We use the criterion from eqn.~\ref{eqn:trapped} to derive the location and boundaries of the \lq\lq capture region" introduced in \S\ref{sec:Underlying}.  Here we show an example for the derivation for the case of a disk with a flat rotation curve.  The maximum and minimum values for the effective potential at the radius of corotation can be expressed as,
\begin{equation}\label{eqn:PhieffMax}
\Phi_{eff}(R_{CR},\phi_{max}) = h_{CR}+|\Phi_s|_{CR}
\end{equation}
and
\begin{equation}\label{eqn:PhieffMin}
\Phi_{eff}(R_{CR},\phi_{min}) = h_{CR}-|\Phi_s|_{CR} .
\end{equation}
The Jacobi integral (eqn~\ref{eqn:EJ}) for a star with zero random energy in a disk with a flat rotation curve can be explicitly written,
\begin{equation}\label{eqn:EJflat}
E_{J,flat} = \dfrac{1}{2}\Omega_p^2(R_{CR}-R)^2 + \Phi(R,\phi)-\dfrac{1}{2}\Omega_p^2R^2 ,
\end{equation} 
where the first term ($E_{rot}=\frac{1}{2}v_{rot}^2=\frac{1}{2}\Omega_p^2(R_{CR}-R)^2$) is an expression for the energy associated with the circular orbital velocity of a star in the rotating frame at some radial distance from corotation.  Eqn~\ref{eqn:Contopoulos78} describes the criterion for stars with zero random energy to be captured in trapped orbits.  We can therefore use eqn.~\ref{eqn:EJflat} to re-express eqn~\ref{eqn:Contopoulos78} as,
\begin{equation}\label{eqn:CaptureRegion}
\Phi_{eff}(R_{CR},\phi_{min}) < \dfrac{1}{2}\Omega_p^2 R_{CR}^2 - \Omega_p^2 R R_{CR} +\Phi(R,\phi) \leq \Phi_{eff}(R_{CR},\phi_{max}) .
\end{equation}
The coordinate-space solutions to the following equation
\begin{equation}\label{eqn:CaptureRegionBoundary}
\Phi_{eff}(R_{CR},\phi_{min}) = \dfrac{1}{2}\Omega_p^2 R_{CR}^2 - \Omega_p^2 R R_{CR} +\Phi(R,\phi) 
\end{equation}
define the boundary to the region of the disk wherein a star with zero random energy will be captured in a trapped orbit.  This expression describes the \lq\lq capture region" introduced in \S\ref{sec:Underlying}, which is shaded grey in fig.~\ref{fig:CR8_PhiEff}.  The value of the difference $\Phi_{eff}(R_{CR},\phi_{max})-\Phi_{eff}(R_{CR},\phi_{min})$ is set by the amplitude of the spiral potential (via eqn.~\ref{eqn:EffectivePotential}) at corotation.  Higher amplitude spirals will, therefore, lead to a larger area for the capture region.  We investigate how the area of the capture region affects the efficiency of radial migration for a population of stars in later papers in this series.

\subsection{Capture Criterion: Stars with Non-Zero Random Orbital  Energy}\label{sec:DerivationWRanMotions}

Stars always have finite random orbital energy.  It is well established that a star's orbital angular momentum and random orbital energy can be altered together when the star encounters a fluctuation in the underlying disk potential away from corotation, particularly at the Lindblad resonances \citep[eg.][]{SS53, BW67, Wielen77, CS85, SB02}.  We will henceforth refer to any event that changes both a star's random orbital energy and its orbital angular momentum as a \lq\lq scattering" event.\footnote{It should be understood that our use of the term \lq\lq scattering" is distinct from the term \lq\lq resonant scattering" used by \cite{SB02} in reference to the changes in orbital angular momentum at corotation with no associated heating.  It is also distinct from the redirection of random motions without heating as is expected from interactions with corotating GMCs \citep{SS53,Lacey84,Sellwood13} or transient spirals \citep{CS85}.}  (We give a detailed discussion of scattering and the consequences of scattering for stars in trapped orbits in \S\ref{sec:Scattering}.)  While stars are, in general, born on nearly circular orbits, scattering events lead to non-circular orbits.  Therefore, the above capture criterion (eqn.~\ref{eqn:trapped}) has limited utility in a disk galaxy and it is necessary to derive a capture criterion for stars on orbits that have random energy. 

Eqn.~\ref{eqn:EJexpanded} can be re-written in the standard quadratic form: 
\begin{equation}
0=A J_\phi^2 +B J_\phi + C 
\end{equation}
by making the following substitutions,
\begin{equation}
\begin{array}{cl}
A&=c_{CR}\\
B&=2b_{CR} J_R\\
C&=-E_J+h_{CR}+|\Phi_s|_{CR} \cos(m \phi_1)+\kappa_{CR}J_R+a_{CR}J_R^2 .
\end{array}
\end{equation}

Real solutions for $J_\phi$ exist when 
\begin{equation}\label{eqn:RealRequirements}
{B^2-4AC} \geq 0
\end{equation} 
is satisfied.  Provided that the disk is not in solid body rotation, $A = c_{CR}$ is inherently negative so that $-c_{CR}= |c_{CR}|$ and we can rewrite eqn.~\ref{eqn:RealRequirements} to give:
\begin{equation}
E_J-h_{CR}-\kappa_{CR} J_R-\left(a_{CR}+\dfrac{b_{CR}^2}{|c_{CR}|}\right) J_R^2 \leq |\Phi_s|_{CR}\cos(m\phi_1) .
\end{equation}

Real solutions for $J_\phi$ exist for all values for $\phi_1$ - and hence the star circulates about the galactic centre in the rotating frame - when,
\begin{equation}\label{eqn:Lambda_nc}
\Lambda_{nc} \equiv \dfrac{1}{|\Phi_s|_{CR}} \left[ E_J-h_{CR}-\kappa_{CR} J_R-\left(a_{CR}-\dfrac{b_{CR}^2}{c_{CR}}\right) J_R^2\right] \leq -1 ,
\end{equation}
where the subscript \lq\lq nc" indicates that $\Lambda_{nc}$ sets the criterion for stars in non-circular orbits in the unperturbed potential.

Solutions are real for only a restricted range of values for $\phi_1$ - and hence the orbit oscillates in azimuthal angle - for $\Lambda_{nc}$ in the range 
\begin{equation}\label{eqn:trappedMe}
-1 < \Lambda_{nc}  \leq 1 .
\end{equation}
A star that meets this criterion (eqn.~\ref{eqn:trappedMe}) will be captured in a trapped orbit librating around the local maximum in the effective potential (e.g. $L_4$).  Note that when $J_R=0$, the quantity $\Lambda_{nc}=\Lambda_c$.  

The value of $\Lambda_{nc}$ is a time-dependent quantity (in contrast to $\Lambda_c$ which is time-independent) since both $J_R$ and $L_z$ are time-dependent (and related through the Jacobi integral) in a non-axisymmetric potential.  A star that meets the criterion in eqn.~\ref{eqn:trappedMe} (and is therefore in a trapped orbit) will experience changes in its value for $J_R$ as its orbital angular momentum oscillates (see eqn.~\ref{eqn:SB02JR}), unless the star also has an instantaneous orbital angular frequency equal to the pattern speed of the perturbation.  Consequently, a star that initially meets the capture criterion in eqn.~\ref{eqn:trappedMe} may not indefinitely continue to do so and could begin to orbit around the galactic centre in the rotating frame. We discuss how the time-dependences of radial action and of angular momentum affect trapped orbits in \S\ref{sec:Scattering}.

\subsection{Special Case: Power-Law Potential}\label{sec:CaptureWithEpi}

Consider the special case of a star moving in a 2D plane under the influence of an unperturbed potential described by a spherical power law \citep[see also][Appx.~B]{Dehnen99} (an approximation for the multicomponent
galactic system - halo, bulge and disk), plus an imposed spiral pattern
\begin{equation}\label{eqn:BetaPotentials}
\Phi_0 = 
   \begin{cases}
   \Phi_{00} \left( \frac{r}{r_p} \right)^{2-\beta} & \text{if } \beta\ne 2\\
   v_c^2 \ln(r/r_p) & \text{if } \beta = 2
   \end{cases} 
\end{equation}
where $r_p$ is the scale length of the potential, $\Phi_{00}$ is a constant, and the index $\beta$ (related to the index used by \cite{Dehnen99} - here denoted $\beta_{Dehnen}$ - by $\beta = 2(1-\beta_{Dehnen})$) can take a value between $0$ (corresponds to solid body rotation) and $3$ (Keplerian motion).  The unperturbed circular frequencies in the plane of the disk vary with cylindrical coordinate $R$ as
\begin{equation}
\Omega_c \propto R^{-\beta/2} .
\end{equation}
The guiding center radius is
\begin{equation}\label{eqn:GuidingCenter}
R_L=R \dfrac{v_\phi}{v_c}, 
\end{equation}
where $v_\phi$ is the instantaneous azimuthal velocity of the star in the inertial frame.  The epicyclic frequency of radial oscillation, $\kappa$ (eqn.~\ref{eqn:kappa}), can now be expressed in terms of the parameter $\beta$ and the circular orbital frequency:
\begin{equation}\label{eqn:kappaOmega}
\kappa^2(R_L)= (4-\beta) \Omega_c^2(R_L) .
\end{equation}
With these approximations, eqns.~\ref{eqn:coeffs} can be expressed in terms of $\beta$ for a given normalisation.

The equations leading to the capture criterion may now be expressed in terms of $\beta$, after substituting for $\kappa$.  In particular, $a_{CR}$ in eqn.~\ref{eqn:Lambda_nc} reduces to:
\begin{equation}
a_{CR}=
  \begin{cases}
  \dfrac{v_c(R_{CR})^4}{R_{CR}^6} \dfrac{(4-\beta)(2-\beta)}{16}[(1-\beta)(\beta)(1+\beta)+60-\frac{5}{3}(4-\beta)(2-\beta)(3+\beta)^2] & \text{if } \beta\ne 2\\
  -\dfrac{11 v_c^4}{3 R_{CR}^6} & \text{if } \beta = 2 ,
  \end{cases}
\end{equation}
where the circular velocity at radial coordinate $R$ is given by $v_c(R)\equiv \sqrt{R\Phi'}$ for $\beta\neq 2$ and $v_c$ is constant for $\beta=2$.

The ratio $-b_{CR}^2/c_{CR}$ in eqn.~\ref{eqn:Lambda_nc} can be expressed as
\begin{equation}
-\dfrac{b_{CR}^2}{c_{CR}	}=\dfrac{\beta}{2R_{CR}^2} .
\end{equation}

\begin{figure}
\begin{center}
\includegraphics[scale=1]{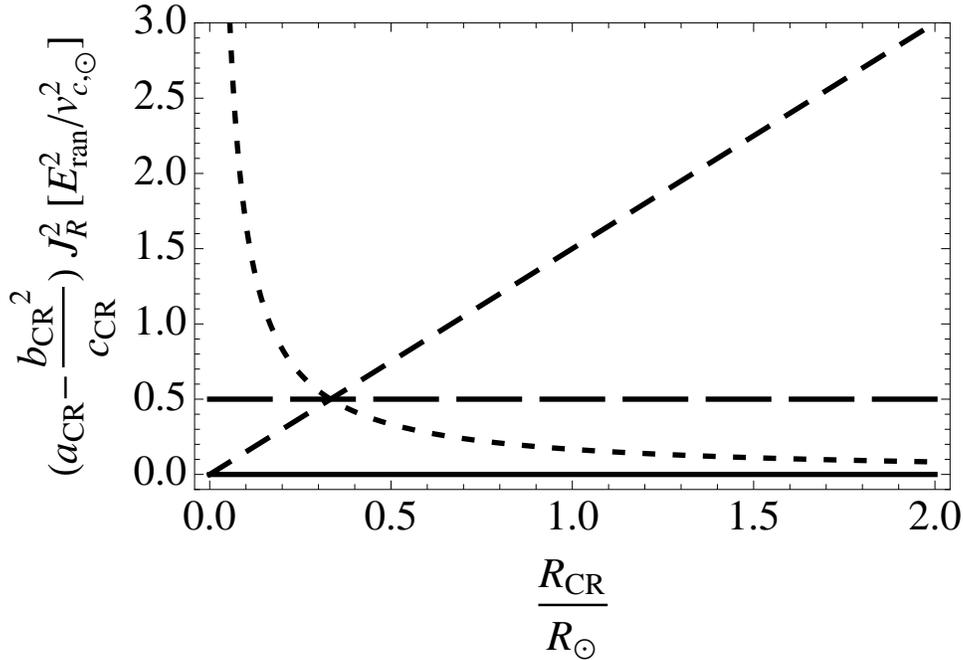}
\caption{Value for $\left( a_{CR}-b_{CR}^2/c_{CR}\right)\,J_R^2$ at $R_{CR}/R_\odot$ for $\beta=0$ (solid; solid body rotation),$1$ (short-dash) $2$ (long-dash; flat rotation curve), and $3$ (medium-dash; Keplerian motion), normalised so that $v_{c,\odot}=v_c(R_\odot)=220$~km~s$^{-1}$, where $R_\odot=8$~kpc.}
\label{fig:CaptureBeta}
\end{center}
\end{figure}

Figure~\ref{fig:CaptureBeta} shows the value of the last term in eqn.~\ref{eqn:Lambda_nc}, $\left(a_{CR}-b_{CR}^2/c_{CR} \right)J_R^2$, as a function of $R_{CR}$, adopting the normalisation so that $v_{c,\odot}=220$~km~s$^{-1}$ at $R_\odot=8$~kpc for all rotation curves and using eqn.~\ref{eqn:SB02JR} to express $J_R$ in terms of random orbital energy in the epicyclic approximation.  We use eqn.~\ref{eqn:SB02JR} to express $\left(a_{CR}-b_{CR}^2/c_{CR} \right)J_R^2$ in units of energy given by $[E_{ran}^2/v_{c,\odot}^2]$.  For $R_{CR}/R_\odot \gtrsim 0.1$, the value of $\left(a_{CR}-b_{CR}^2/c_{CR} \right)J_R^2$ is of order unity for all allowed values of $\beta$.  In comparison, the third term in eqn.~\ref{eqn:Lambda_nc} (i.e.~$\kappa_{CR} J_R$) is equal to $v_{c,\odot}^2/E_{ran}$, in units of $[E_{ran}^2/v_{c,\odot}^2]$.\footnote{From eqn.~\ref{eqn:SB02JR}, $\kappa_{CR} J_R = E_{ran}=v_{c,\odot}^2/E_{ran}\,[E_{ran}^2/v_{c,\odot}^2].$}   This ratio is large for the epicyclic approximation to be valid.  For example, if one assumes that the disk has $v_{c,\odot}$ of order $10^2$~km~s$^{-1}$ and random velocity associated with the random energy (i.e.~$v_{ran}=\sqrt{2 E_{ran}}$) of order $10$~km~s$^{-1}$, then $\kappa_{CR} J_R$ is of order $10^3$, while $\left(a_{CR}-b_{CR}^2/c_{CR} \right)J_R^2$ is of order unity.  We therefore omit the last term ($\left(a_{CR}-b_{CR}^2/c_{CR} \right)J_R^2$) from eqn.~\ref{eqn:Lambda_nc} and write,
\begin{equation}\label{eqn:Lambda_ncbeta}
\Lambda_{nc,\beta}  = \dfrac{1}{|\Phi_s|_{CR} } \left[ E_J-h_{CR}-\kappa_{CR} J_R \right] .
\end{equation}
The subscript, $\beta$, signifies that eqn.~\ref{eqn:Lambda_ncbeta} is a good approximation within an underlying potential set by eqn.~\ref{eqn:BetaPotentials} for $0\leq\beta\leq 3$.

Combining eqns.~\ref{eqn:Lambda_ncbeta} and \ref{eqn:trappedMe} we find the capture criterion for a star near the corotation resonance of a spiral that has pattern speed $\Omega_p$ and amplitude at corotation, $|\Phi_s|_{CR}$, to be,
\begin{equation}\label{eqn:trappedMeFlat}
-|\Phi_s|_{CR}  < E_J-h_{CR}-\kappa_{CR} J_R \leq |\Phi_s|_{CR}  .
\end{equation}

Substituting $E_{ran}/\kappa$ for the radial action (as above) gives the following expression:
\begin{equation}\label{eqn:trappedMeFlatEpi}
-|\Phi_s|_{CR}  \leq E_J -h_{CR} -\left(\dfrac{R_L(t)}{R_{CR}}\right)^{\beta/2}E_{ran}(t)  \leq |\Phi_s|_{CR}  ,
\end{equation}
where we have explicitly shown the time dependent quantities in the non-axisymmetric potential.  

Eqn.~\ref{eqn:trappedMeFlatEpi} gives the criterion for a star to be captured in a trapped orbit, in terms of orbital energy and orbital angular momentum (via $R_L(t)$), for all disk stars on orbits for which the epicyclic approximation holds.  We expect the criterion in eqn.~\ref{eqn:trappedMeFlatEpi} to break down for highly eccentric orbits or for excursions in radius beyond the validity of the approximation of constant density in the underlying matter distribution.  Rearranged and in a potential with a flat circular velocity (rotation) curve ($\beta=2$),
\begin{equation}\label{eqn:Lambda_nc2}
\Lambda_{nc,2}(t) \equiv \Lambda_c - \left(\dfrac{R_L(t)}{R_{CR}}\right) \left(\dfrac{E_{ran}(t)}{|\Phi_s|_{CR} }\right) ,
\end{equation}
The capture criterion for stars in orbits that are not highly eccentric is, in this case, to a very good approximation, 
\begin{equation}\label{eqn:trappedMeFlatEpiLz}
-1\, < \Lambda_{nc,2}(t) \leq\, 1 .
\end{equation}
As expected, $\Lambda_{nc,2}\rightarrow\Lambda_c$, the criterion for stars with zero radial action, in the limit that $E_{ran}\rightarrow 0$.  The physical parameters that determine whether or not a star is captured in  a trapped orbit are embedded in eqn.~\ref{eqn:trappedMeFlatEpiLz}.

We showed in \S\ref{sec:CaptureRegion} that a star with zero random energy must have physical coordinates $(R,\phi)$ in the capture region (i.e. the grey region in fig.~\ref{fig:CR8_PhiEff}) to be in a trapped orbit.  Solutions to eqn.~\ref{eqn:CaptureRegionBoundary} define the size, location, and shape of the capture region in a flat rotation curve.  We now explore the significance of this capture region for stars with non-zero random orbital energy.

\begin{figure}
\begin{center}
\includegraphics[scale=1]{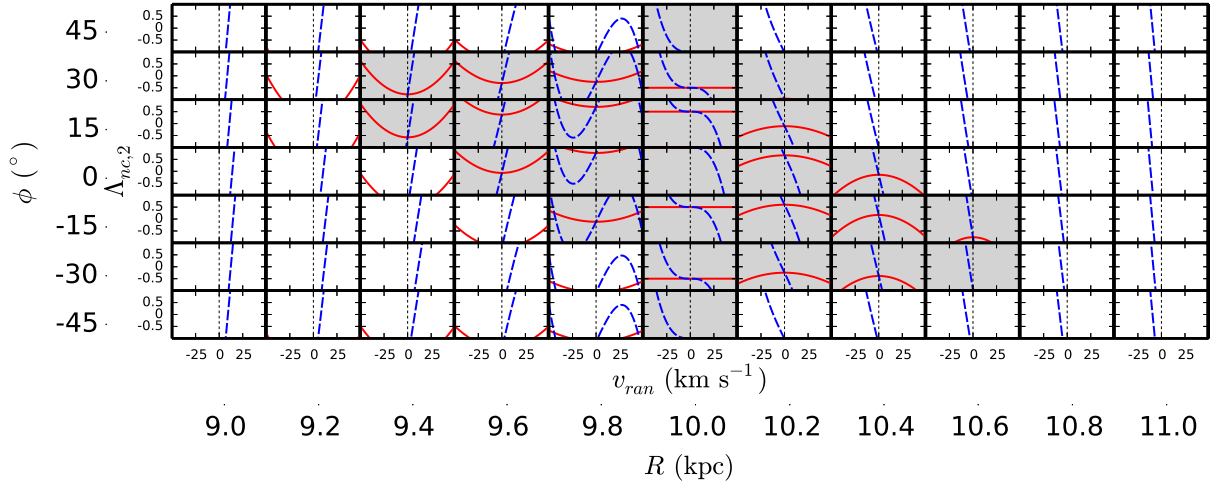}
\caption{Each panel shows the value of the random velocity in the inertial frame ($v_{ran}$) on the x-axis and its associated value of $\Lambda_{nc,2}$ on the y-axis, (within the range $-1<\Lambda_{nc,2}\leq 1$, which satisfies the capture criterion), for a star at a given coordinate.  The coordinate corresponding with each panel falls in the range $9$~kpc$\leq R\,\leq 11$~kpc in $0.2$~kpc intervals (horizontal direction) and $-45^\circ\leq \phi \leq 45^\circ$ in $15^\circ$ intervals (vertical direction).  The coordinate grid is labelled on the exterior of the plot.  We have assumed a spiral with $R_{CR}=10$~kpc, $\theta=10^\circ$, and $\epsilon_\Sigma=0.2$, in an underlying potential with a flat rotation curve and circular velocity $v_c=220$~km~s$^{-1}$.   Solid (red) curves show the value of $\Lambda_{nc,2}$ corresponding to $\mathbf{v}_{ran}=|v_{ran}|\mathbf{\hat{R}}$, and the dashed (blue) curves are for $\mathbf{v}_{ran}=|v_{ran}|\boldsymbol{\hat{\phi}}$.  The capture region is defined as the locus within which a star with zero random motion in the inertial frame ($v_{ran}=0$ marked by a vertical, dotted line) satisfies the capture criterion.  Panels for coordinates that are within the capture region have a shaded background.  For a star with coordinates within the capture region, there is a large range in radial random motion that gives rise to orbits that satisfy the capture criterion (solid, red), suggesting that random orbital energy alone is not a significant determining factor for trapped orbits.  The range of random motion in the azimuthal direction - associated with orbital angular momentum as well as random orbital energy - for a star to satisfy the capture criterion (dashed, blue) is relatively small, leading to the conclusion that whether or not a star is captured in a trapped orbit is sensitive to the value of that star's orbital angular momentum.  A star with coordinates that is outside the capture region may still satisfy the capture criterion.  For example, the panel with $R=9.2$~kpc and $\phi=30^\circ$ illustrates that over a narrow range of random velocities in either direction $|\Lambda_{nc,2}|<1$.
}
\label{fig:CaptureRegionWithLambda}
\end{center}
\end{figure}

The direction of $v_{ran}$ for a star at a given coordinate is important for determining whether or not that star meets the capture criterion, $-1<\Lambda_{nc,2}\leq 1$.  Figure~\ref{fig:CaptureRegionWithLambda} illustrates values of a star's random velocity (in the inertial frame) that satisfy the capture criterion for a spiral with $R_{CR}=10$~kpc, $\theta=10^\circ$, and $\epsilon_\Sigma=0.2$ in an underlying potential with a flat rotation curve and $v_c=220$~km~s$^{-1}$.  Shaded panels indicate coordinates within the capture region (note that a star with $v_{ran}=0$ satisfies the capture criterion only in the capture region).  The curves in Fig.~\ref{fig:CaptureRegionWithLambda} show values of $\Lambda_{nc,2}$, within the range  $-1<\Lambda_{nc,2}\leq 1$, for $v_{ran}$ entirely in the radial direction (solid, red) and entirely in the azimuthal direction (dashed, blue).  Note that in the capture region, the red curve spans a broad range of random radial velocities, indicating that the value of $\Lambda_{nc,2}$ is not sensitive to radial random motion.  This contrasts with the dashed, blue curve, which satisfies the capture criterion for only a restricted range of velocities, illustrating that $\Lambda_{nc,2}$ is sensitive to azimuthal random motion.  The direction of the random velocity is important because the azimuthal velocity determines the angular momentum of the star, and therefore its guiding centre radius, $R_L$ (eqn.~\ref{eqn:GuidingCenter}), whereas a star with random motion solely in the radial direction will have $R=R_L$.

Consider the situation of two stars (Star A and Star B) that have the same instantaneous position ($\mathbf{x}$) and instantaneous angular momentum ($R_{L,A}=R_{L,B}$), but have different values for their instantaneous random energy ($E_{ran}$).  Star A has $E_{ran,A}=0$ and Star B has some finite random energy, $E_{ran,B}$. The value of $E_J$ for Star B will therefore be greater than that of Star A ($E_{J,B}>E_{J,A}$).  All the instantaneous random energy must be in the form of radial motion at that time and therefore the rotational ($E_{rot}$) and random (radial) components of $\dot{\mathbf{x}}^2$ in eqn.~\ref{eqn:EJ} are orthogonal, and $E_{J,A}=E_{J,B}-E_{ran,B}$.  With the capture region defined as in eqn.~\ref{eqn:EJflat} ($\Phi_{eff}(R_{CR},\phi_{min})=E_{rot}(R)+\Phi_{eff}(R,\phi)$), and given that the velocities leading to $E_{rot}$ and $E_{ran}$ are orthogonal, the capture criterion (eqn.~\ref{eqn:trappedMeFlatEpi}) can be written as:
\begin{equation}
\Phi_{eff}(R_{CR},\phi_{min}) < 
E_{rot}(R) +\Phi_{eff}(R,\phi) +\left(1-\dfrac{R_L(t)}{R_{CR}}\right)^{\beta/2}E_{ran}(t)
\leq \Phi_{eff}(R_{CR},\phi_{max}).
\end{equation}
Should Star~A be within the capture region at the time under consideration then $R_L\approx R_{CR}$.  In that case, even though Star~B may have a different value for $E_{ran}$, given that $R_{L,B}=R_{L,A}$, the value of the random energy is of little consequence as to whether or not the star meets the capture criterion.\footnote{For $R_L$ away from corotation, other gravitational processes, like interactions with the Lindblad resonances, become important.  We discuss this in more detail in \S\ref{sec:Scattering}}  For example, the solid, red curve in the panel of Fig.~\ref{fig:CaptureRegionWithLambda} where $R=10.2$~kpc and $\phi=-15^\circ$ shows that the value of $\Lambda_{nc,2}$ has little dependence on the value of the random energy.  

The case of a star with random velocity in the azimuthal direction is more complex.  The rotational ($v_{rot}$) and random ($v_{ran}$) velocities in the rotating frame are not orthogonal, and therefore the random orbital energy and the angular momentum are not independent.  The dashed, blue curve in Fig.~\ref{fig:CaptureRegionWithLambda} shows the value of $\Lambda_{nc,2}$ in the limiting case of a star particle that has its random motion entirely in the azimuthal direction.  In every panel in Fig.~\ref{fig:CaptureRegionWithLambda}, it is clear that the value of $\Lambda_{nc,2}$ has a strong dependence on the value of the random velocity in the azimuthal direction.  In combination with the above discussion on radial motion, one can conclude that whether or not a star is in a trapped orbit is primarily determined by its orbital angular momentum ($R_L(t)$), and less so by $E_{ran}(t)$.

Finally, when a single star is scattered (i.e.~it has changes in angular momentum with associated changes in random energy by eqns.~\ref{eqn:SB02JR}), the value of $\Lambda_{nc,2}(t)$ may change to such a degree that the star no longer meets the capture criterion (eqn.~\ref{eqn:trappedMeFlatEpiLz}).  This arises since a star's value of $E_J$ (and therefore $\Lambda_c$) is conserved, while $E_{ran}(t)$ and $R_L(t)$ are time dependent.  We discuss this in more detail in \S\ref{sec:Scattering}.

In the next section we carry out numerical tests of the predictive power of eqn.~\ref{eqn:trappedMeFlatEpiLz}, in a 2D disk with an imposed spiral perturbation.

\section{Numerical Exploration}\label{sec:NumericalExploration}

\subsection{The Approach}\label{sec:NumericalExplorationModel}

We use an orbital integrator\footnote{We use a $2^{nd}$ order leapfrog orbital integrator.} to follow the orbits of test particles in an underlying potential.  We follow each particle for $2\times10^9$~yr using $10^2$~yr fixed time-steps.   We verified that differences of $1$~km~s$^{-1}$ in the initial orbital velocity did not lead to largely divergent orbital paths in the underlying axisymmetric potential; we therefore conclude that our choice for the length of the time-step is sufficient for this analysis.  The mean fractional deviation in the value of the Jacobi integral per time-step is of order $\sim 10^{-5}$. 

We adopt the underlying potential given by eqn.~\ref{eqn:BetaPotentials} with $\beta=2$ and circular speed chosen to be $v_c=220$~km~s$^{-1}$, thus providing a flat rotation curve in the plane of the disk.  Our chosen value for the potential scale length, $R_p=1$~kpc, renders changes in $\Phi_0(R)$ for stars in trapped orbits (usually moving a radial distance $<1$~kpc around $R_{CR}$) to be $|\Delta \Phi_0|/|\Phi_0|<10\%$ between $4<R_{CR}<15$~kpc.  

We adopt an exponential radial surface density for the disk, $\Sigma(R)=\Sigma_0 e^{-R/R_d}$, with the disk scale length $R_d=2.5$~kpc and the disk surface density normalised so that $\Sigma(R=8$~kpc$)=50$~M$_\odot$~pc$^{-2}$, mimicking the solar neighbourhood. 

We superimpose a perturbation to the underlying potential corresponding to an $m$-armed Lin-Shu spiral density wave 
\citep*{LS64,LYS69,BT08} with pattern speed, $\Omega_p$:
\begin{equation}
\Phi_1(R,\phi,t)=\Phi_s(R) \cos \left[\alpha \,\ln (R/R_{CR}) + m\Omega_pt -m\phi \right]
\end{equation}
in the inertial frame, where $\alpha=m\cot \theta$ with $\theta$ as the (constant) spiral pitch angle (small $\theta$ corresponds to tightly wound arms).  The amplitude of the spiral potential is given by
\begin{equation}\label{eqn:SpiralAmplitude}
\Phi_s(R) = \dfrac{2\pi G\Sigma(R) \epsilon_\Sigma}{k(R)} ,
\end{equation}
where $\Sigma(R)$ is the underlying disk surface density, $\epsilon_\Sigma$ is the fractional amplitude in surface density of the spiral pattern and the radial wave-number, $k(R)$, is given by $k(R)=\alpha/R$ \citep[][eqn 6.7]{BT08}.   Consequently the spiral amplitude ($\Phi_s$) is larger for small wavenumber ($k$) in our model.  Further, the spiral amplitude for fixed pitch angle ($\theta$) peaks at the disk scale length ($R_p$), since $\Sigma(R)/k(R) \propto Re^{-R/R_d}$ for an exponential disk.  Note that the choice for the radius of corotation, fractional amplitude in surface density, spiral wave number, and pitch angle of the spiral pattern therefore affects the the size, shape and location of the capture region (\S\ref{sec:CaptureRegion}).  We will investigate the effects of these choices in a later paper of this series.

For illustrative purposes we adopt a spiral amplitude that is typically higher (e.g. $\theta=25^\circ$ and $\epsilon_\Sigma=0.3$) than estimates from observations of external disk galaxies \citep[e.g.][]{RZ95,SJ98,Ma02}.
Our choice of spiral amplitude causes the capture region to be larger than one would expect from a more modest choice and consequently a larger fraction of stars in the disk will meet the capture criterion.

\subsection{Testing the Capture Criteria}\label{sec:NumericalTestingCaptureCriteria}

We model stars as test particles and compute the orbits assuming a range of initial conditions.  We adopt parameter values for the spiral patterns such that $0.1\leq\epsilon_\Sigma\leq 0.5$, $5^\circ\leq\theta\leq 45^\circ$, $0\leq m\leq 4$ and $4\leq R_{CR}\leq 15$~kpc.  Each test particle is launched with an initial position in the rotating frame $(R(t_0),\phi_1(t_0))\equiv (R_0,\phi_{1,0})$ that is within $5$~kpc of corotation and at a range of azimuthal positions\footnote{We set the azimuthal coordinate $\phi=0$ at the position of the minimum surface density, which is located between the spiral arms}.  Each initial velocity 
\begin{equation}
\mathbf{v}_0 =\mathbf{v}_{ran,0}+\mathbf{v}_c 
=\mathbf{v}_{ran,0}+v_c\boldsymbol{\hat{\phi}} 
=v_{ran,0,R}\mathbf{\hat{R}}+(v_c+v_{ran,0,\phi})\boldsymbol{\hat{\phi}} 
\end{equation}
has a speed in any direction up to $50$~km~s$^{-1}$.

For each set of initial conditions, we test (1) the validity of the appropriate capture criterion (eqn.\ref{eqn:trapped}~or~\ref{eqn:trappedMeFlatEpiLz}) and (2) the importance of orbital angular momentum and random energy to determining whether or not a star is in a trapped orbit.  We do not show the results from our exploration across the entire range of initial conditions, but rather a representative subset, where we have adopted $\theta=25^\circ$, $\epsilon_\Sigma=0.3$ and $m=2$~or~$4$.

The choice of direction for the initial random velocity of a star determines the phase on its epicyclic orbit and thus the coordinate of its guiding centre at $t=0$.  Therefore, the primary difference between launching a star at ($R_0$,$\phi_{1,0}$) with given random energy and initial random velocity ($E_{ran,0}\equiv E_{ran}(t=0)=\frac{1}{2}v_{ran,0}^2$ per unit mass) either entirely in the $\mathbf{\hat{R}}$-direction or entirely in the $\boldsymbol{\hat{\phi}}$-direction is that the former corresponds to an orbit with $R_{L,0}\equiv R_L(t=0)=R_0$, while the latter corresponds to an orbit with $R_{L,0}= R_0 (v_{ran,0,\phi} +v_c)/v_c$ (eqn.~\ref{eqn:GuidingCenter}).

\begin{figure}
\begin{center}
\includegraphics[scale=0.5]{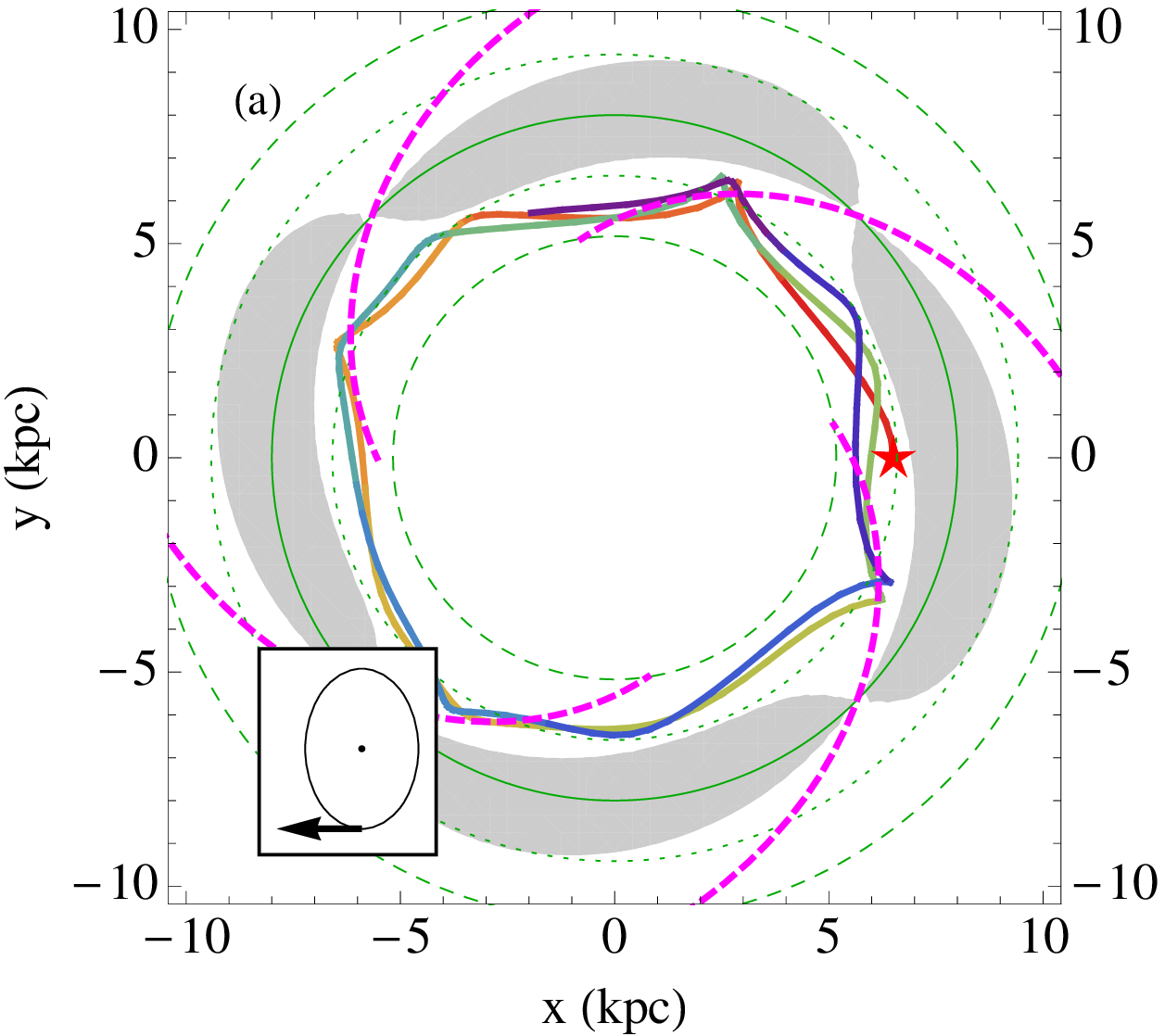}
\includegraphics[scale=1]{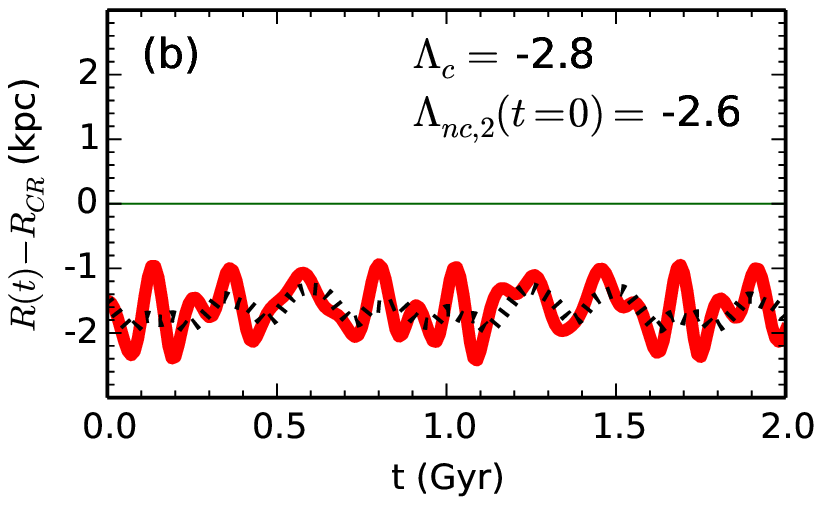}
\caption{Orbital trajectory and properties for a star that is not in a trapped orbit.  The potential is modified from that shown in fig.~\ref{fig:CR8_PhiEff} by setting $R_{CR}=8$~kpc and $\theta=25^\circ$.  The test particle is launched with zero random energy ($|v_{ran,0}|=0$~km~s$^{-1}$) at $(R_0, \phi_{1,0})=(6.5$~kpc$,0)$ (red star).  Panel (a) shows the orbital trajectory in the rotating frame (solid, rainbow) for 2~Gyr ($\gtrsim 2$~orbital periods).  The shaded area shows the capture region and the inset shows the phase of the star's epicyclic orbit at the time of launch.  The solid dark-green line marks the radius of corotation, $R_{CR}$, in both panels.  Panel (b) shows the time-dependent guiding centre radius, $R_L(t)-R_{CR}$ (black, dotted), and radial coordinate $R(t)-R_{CR}$ (solid, red).  The value for $\Lambda_c$ and the initial value for $\Lambda_{nc,2}(t=0)$ are printed in panel (b).}  
\label{fig:CR8_circular}
\end{center}
\end{figure}

Figure~\ref{fig:CR8_circular} shows an orbit for a test particle that does not meet either capture criterion ($\Lambda_c=-2.8$ and $\Lambda_{nc,2}(t=0)=-2.6$, as printed in panel (b)) at the time of launch and is therefore expected not to be in a trapped orbit (\S\ref{sec:Capture}).\footnote{Recall that the capture criteria  (eqns.~\ref{eqn:trapped} \& \ref{eqn:trappedMeFlatEpiLz}) predict that a star is in a trapped orbit only for stars with a value of $\Lambda_c$ or $\Lambda_{nc,2}(t)$ (respectively) between $-1$ and $1$.}  The test particle has initial coordinate $(R_0, \phi_{1,0})=(6.5$~kpc$,0)$ and initial random velocity $|v_{ran,0}|= 0$~km~s$^{-1}$, in an $m=4$ spiral pattern with $R_{CR}=8$~kpc.  This orbit is not circular, despite having zero initial random velocity, since neither angular momentum nor random energy is conserved in a non-axisymmetric potential.  Note that neither $R_0$ nor $R_{L,0}$ is in the capture region and the star orbits around the galactic centre.  Panel (a) shows the orbital path in the rotating frame (solid, rainbow) for a star with initial position marked with a red star.  The solid dark-green curve in both panels indicates the radius of corotation.  The shaded grey area shows the capture region, and the curved lines (thin, magenta) show the location of the spiral arms.  The epicyclic phase at launch is shown in the inset of panel (a) (positioned arbitrarily so as not to obscure the figure).  Panel (b) shows $R_L(t)-R_{CR}$ (dotted, black) and $R(t)-R_{CR}$ (solid, red) as a function of time.  A star that is in a trapped orbit will have a guiding centre radius, $R_L(t)$, that oscillates about the radius of corotation.  

\begin{figure}
\begin{center}
\includegraphics[scale=0.5]{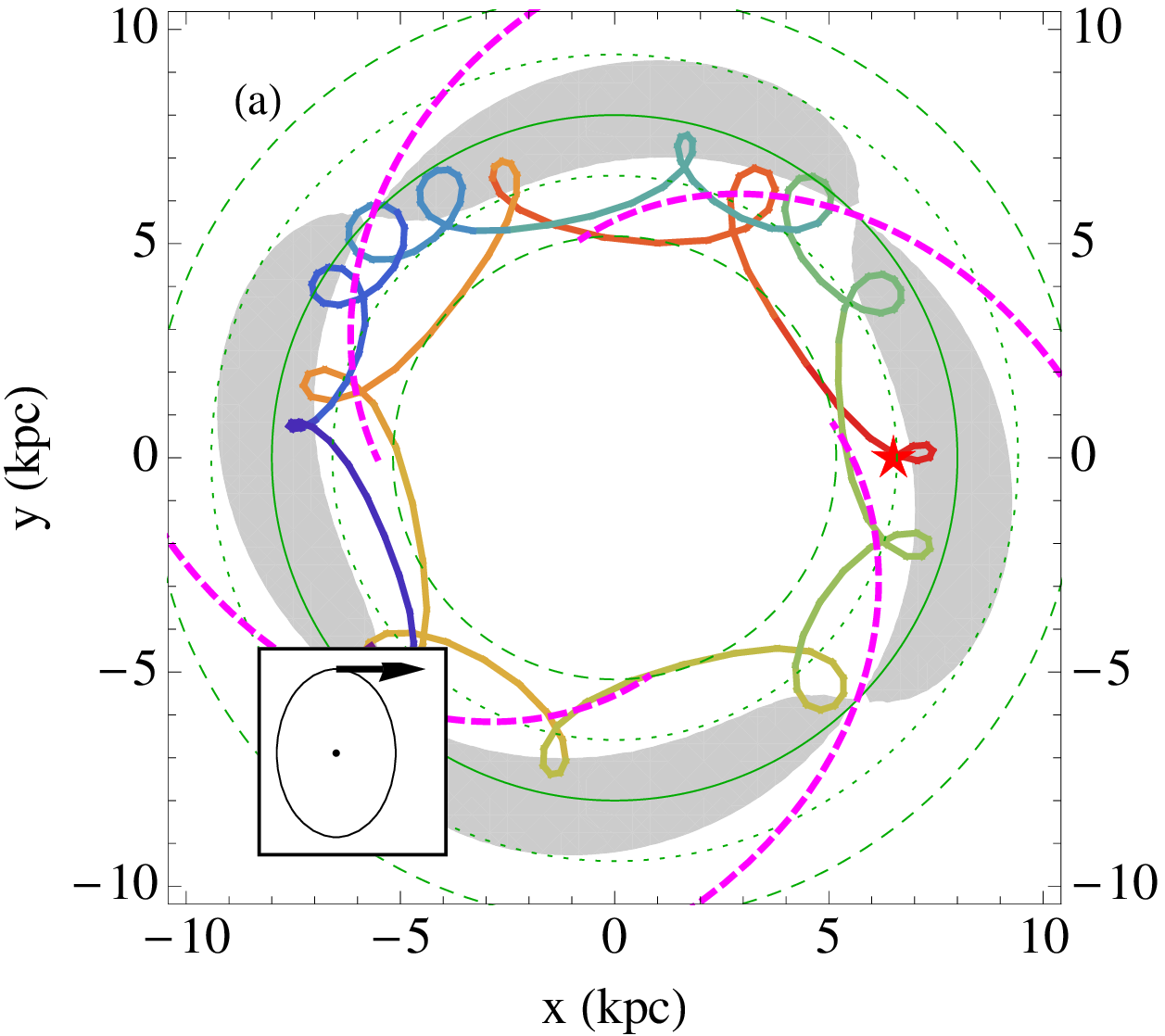}
\includegraphics[scale=1]{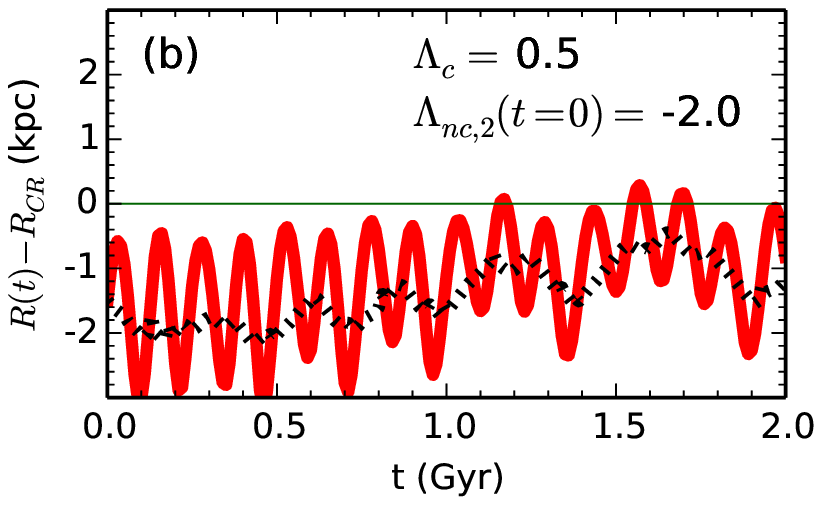}
\caption{The panels and plotted colours and curve patterns have the same meaning have the same as in fig.~\ref{fig:CR8_circular}.  Initial conditions are the same except the initial radial velocity is modified so that $v_{ran,0,R}=50$~km~s$^{-1}$.  This star's orbital trajectory enters the capture region but its guiding centre does not.  As expected from the capture criterion, 
this star is not in a trapped orbit.}
\label{fig:CR8_randomvR}
\end{center}
\end{figure}

Figure~\ref{fig:CR8_randomvR} shows a test particle launched with the same set of initial conditions used to produce the orbit in Fig.~\ref{fig:CR8_circular}, except that the initial random motion in the radial direction is modified so that $v_{ran,0,R}=50$~km~s$^{-1}$.  A test particle in this orbit has the same guiding centre radius as the star in fig.~\ref{fig:CR8_circular}, but its value for $E_{ran,0}$ is higher.  The discussion at the end of \S\ref{sec:CaptureWithEpi} would suggest that this test particle should not be in a trapped orbit, even when its orbital trajectory enters the capture region.  Indeed, panel (a) of fig.~\ref{fig:CR8_randomvR} shows that the test particle orbits the galactic center in the rotating frame, and panel (b) shows that the guiding centre radius does not oscillate about the radius of corotation (both indicating that it is not in a trapped orbit).  The two capture criteria (eqns.~\ref{eqn:trapped}~\&~\ref{eqn:trappedMe}) give different predictions; the capture criterion for a star with zero random energy suggests that the star is in a trapped orbit ($\Lambda_c=0.5$), whereas the capture criterion we derive is not satisfied ($\Lambda_{nc,2}(t=0)=-2.0$), suggesting that the test particle is not in a trapped orbit.  We conclude that the capture criterion derived in \S\ref{sec:DerivationWRanMotions} accurately predicts whether or not this is a trapped orbit.

\begin{figure}
\begin{center}
\includegraphics[scale=0.5]{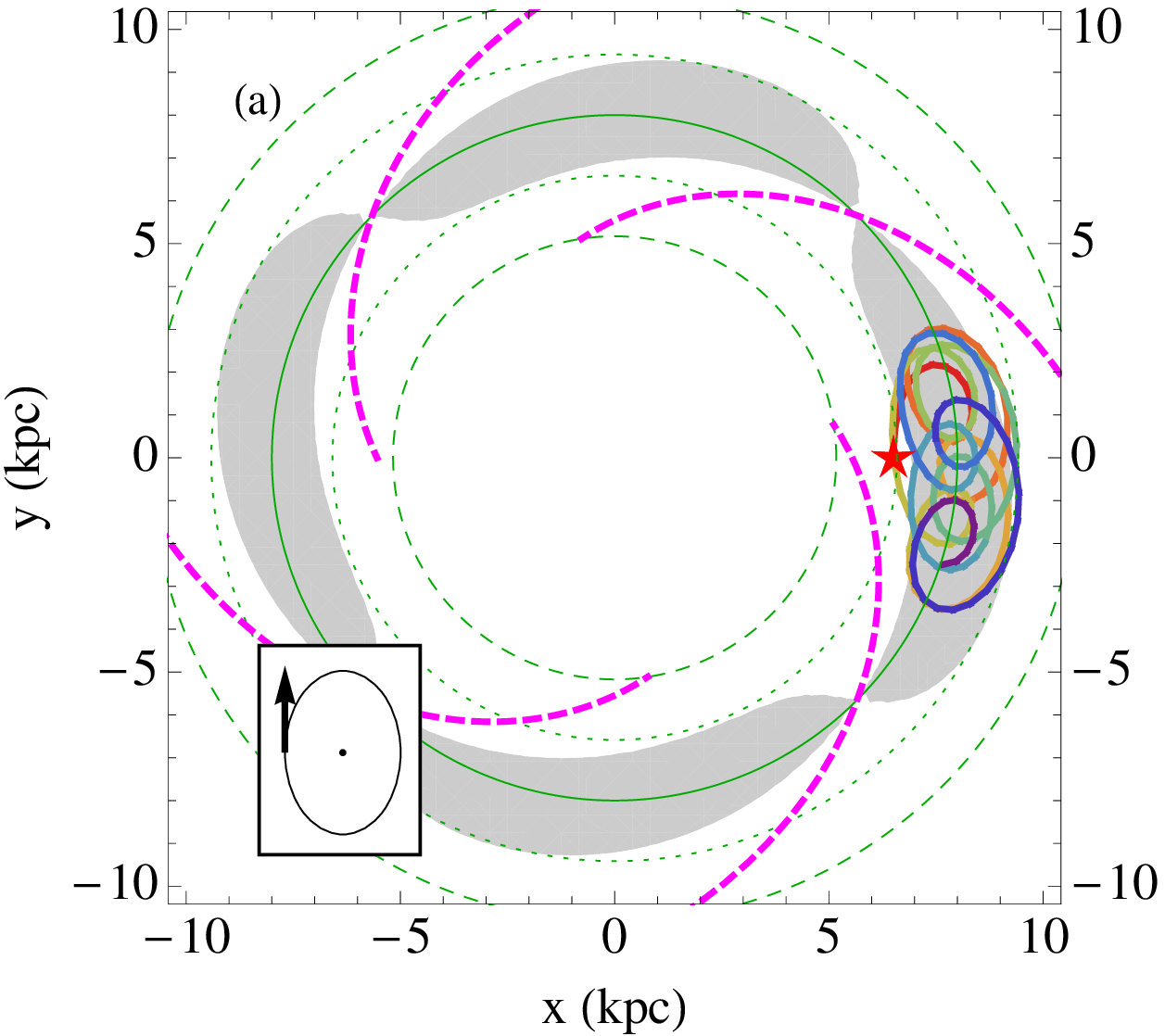}
\includegraphics[scale=1]{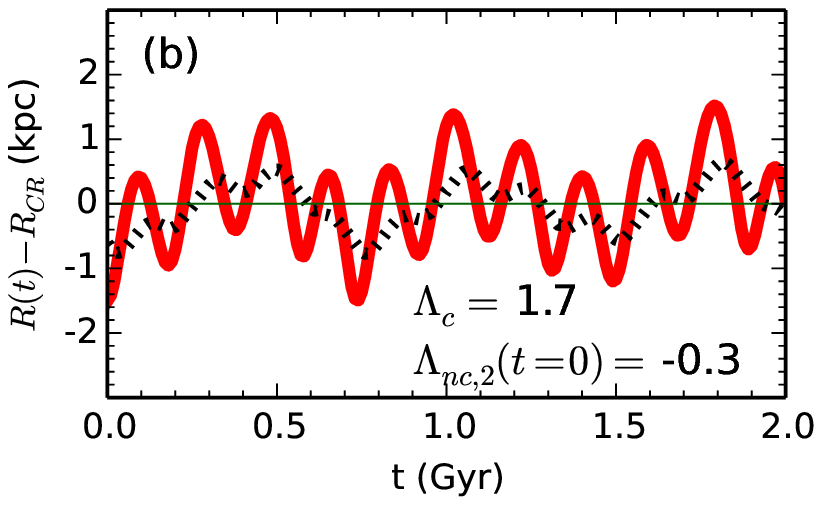}
\caption{The panels and plotted colours and curve patterns have the same meaning as in fig.~\ref{fig:CR8_circular}.  Initial conditions are also the same except the initial azimuthal velocity is modified so that $v_{ran,0,\phi}=30$~km~s$^{-1}$.
The orbital trajectory leaves the capture region, but $R_L$ remains within the capture region.  This star is in a trapped orbit.}
\label{fig:CR8_randomvphi}
\end{center}
\end{figure}

Figure~\ref{fig:CR8_randomvphi} shows a test particle that is launched with the same set of initial conditions used to produce the orbit in fig.~\ref{fig:CR8_circular} except that the initial random motion in the azimuthal direction is modified so that $v_{ran,0,\phi}=30$~km~s$^{-1}$.  A test particle in this orbit has both higher random orbit energy and orbital angular momentum ($R_{L,0}=7.4$~kpc) than the test particle in fig.~\ref{fig:CR8_circular}.  The capture criterion for a star with zero random energy suggests that the star is not in a trapped orbit ($\Lambda_c=1.7$), whereas the capture criterion we derived is satisfied ($\Lambda_{nc,2}(t=0)=-0.3$), leading to the expectation that the test particle is in a trapped orbit.  The figure confirms the prediction given by our capture criterion ($\Lambda_{nc,2}(t=0)$), as the test particle oscillates about the corotation radius and between the spiral arms indicating that it is in a trapped orbit.  Again, the capture criterion derived in \S\ref{sec:DerivationWRanMotions} accurately predict whether or not this is a trapped orbit.  In this case, the guiding center radius is within the capture region, while the orbital trajectory is not confined to it.

\begin{figure}
\begin{center}
\includegraphics[scale=0.5]{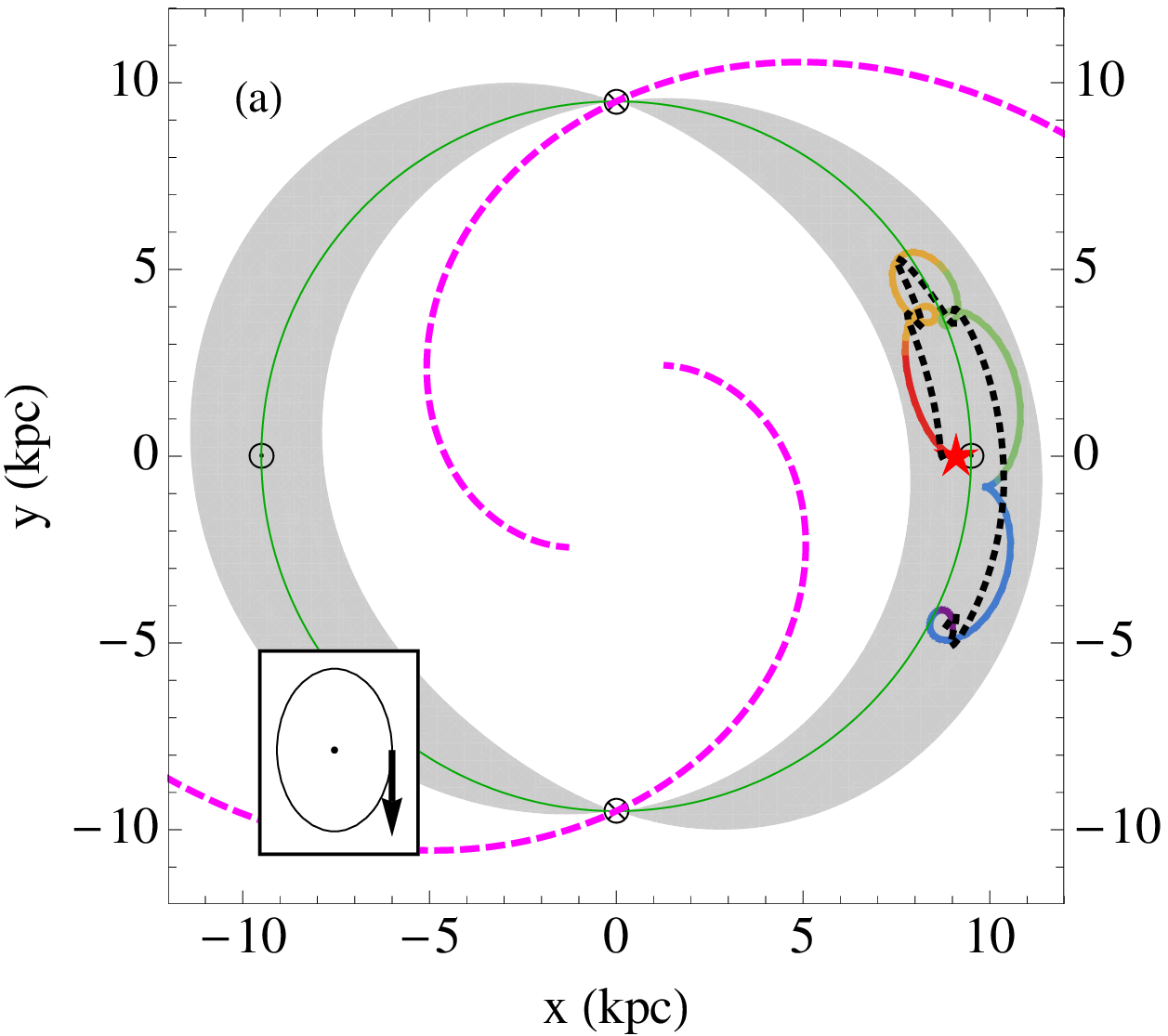}
\includegraphics[scale=1]{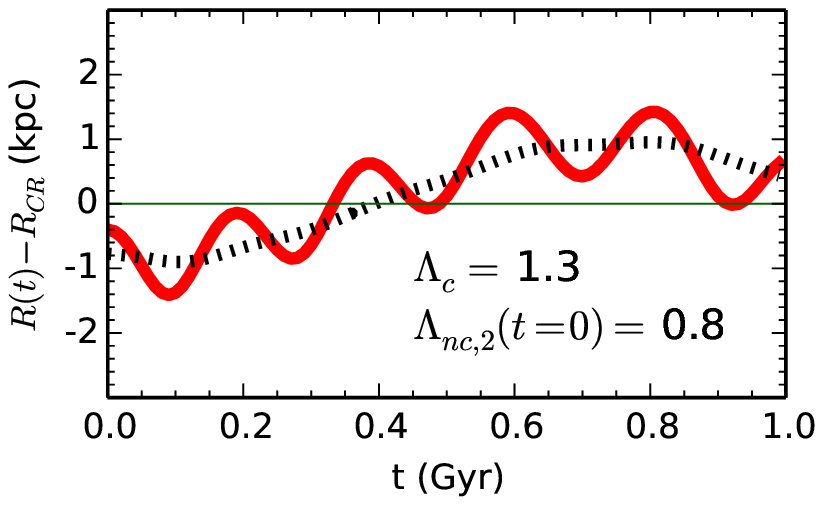}\\
\includegraphics[scale=0.5]{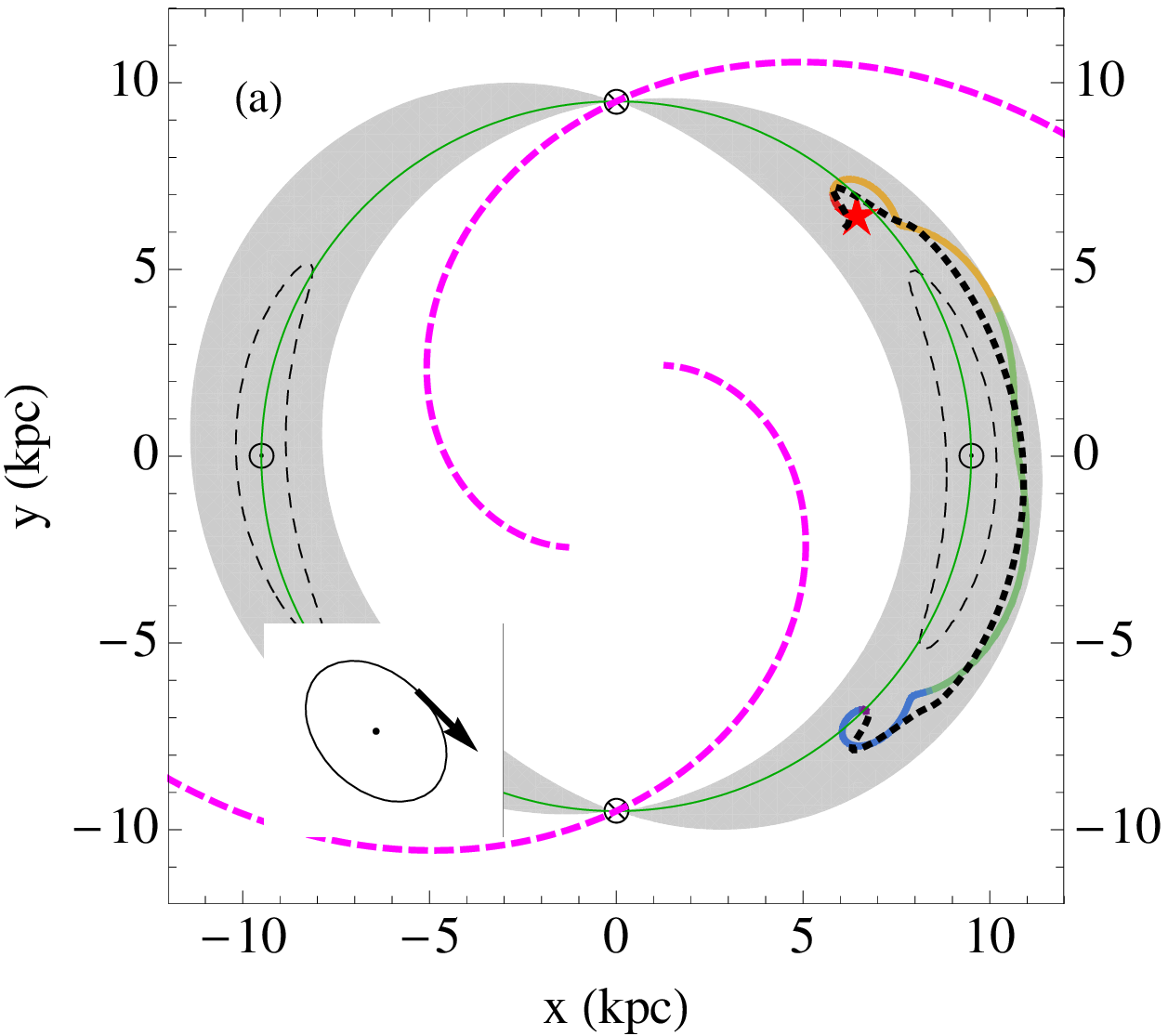}
\includegraphics[scale=1]{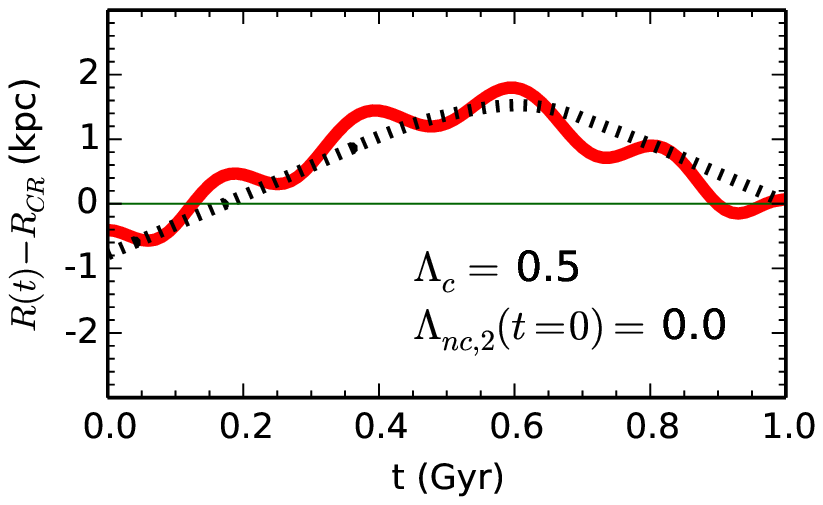}\\
\includegraphics[scale=0.5]{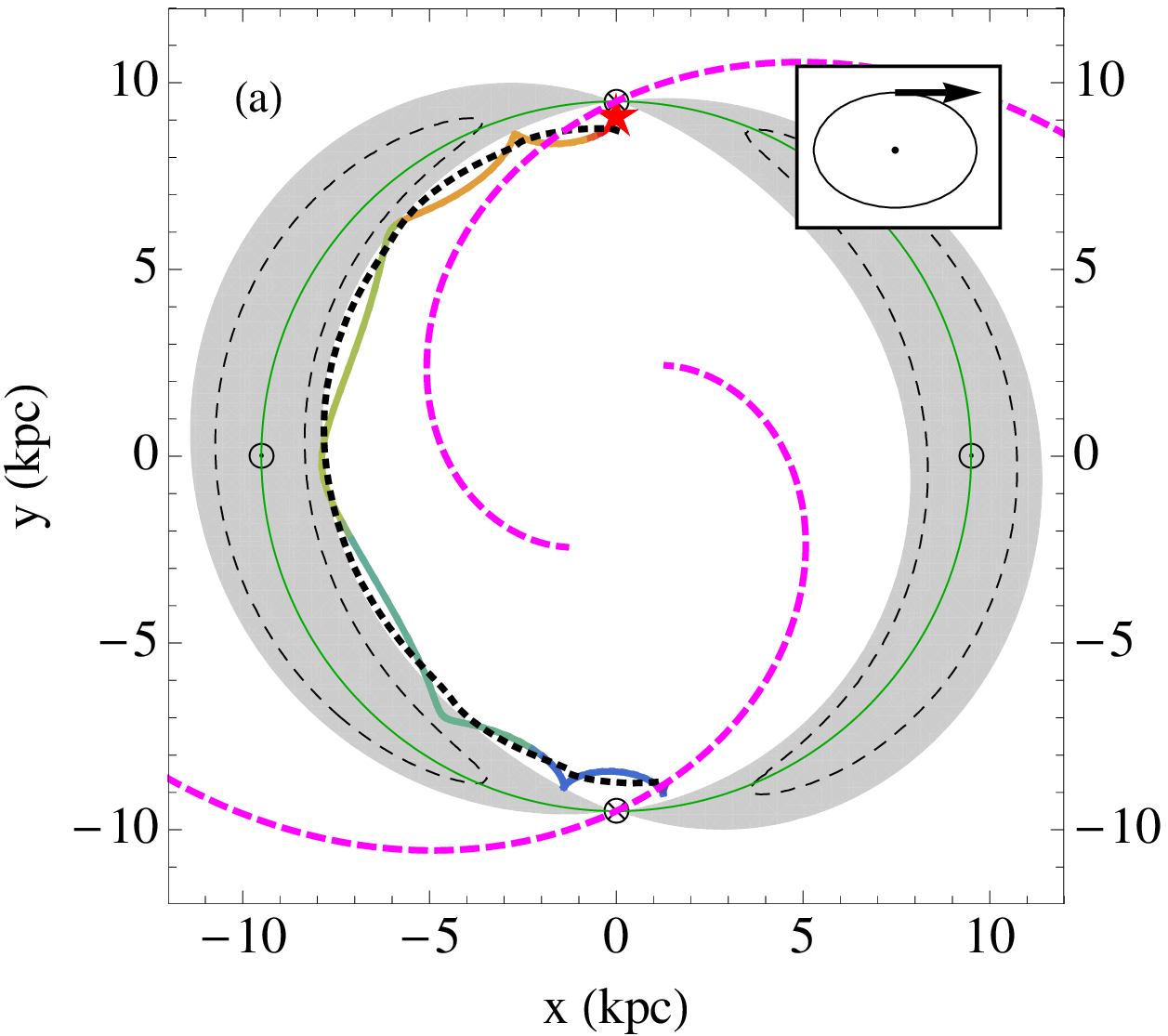}
\includegraphics[scale=1]{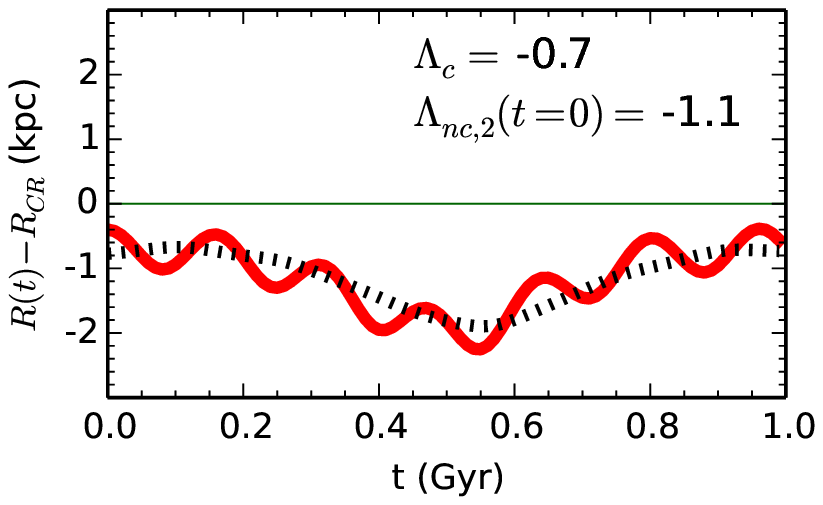}
\caption{The panels and plotted colours and curve patterns have the same meaning as in fig.~\ref{fig:CR8_circular}.  Initial conditions are modified so that $m=2$, $R_{CR}=9.5$, $v_{ran,0,\phi}=-10$~km~s$^{-1}$ and from top to bottom $\phi_{1,0}=\lbrace 0, \pi/4, \pi/2 \rbrace$.  Each particle is launched at $R_0=9.1$~kpc, the apocentre of its epicycle.  The top and middle panels show a star that is in a trapped orbit, and the bottom panels show a star that is not in a trapped orbit.}
\label{fig:phi0_vs_trapped}
\end{center}
\end{figure}

In figure~\ref{fig:phi0_vs_trapped}, we explore how the azimuthal position of a star affects whether or not it is in a trapped orbit, given the same initial radius ($R_0=9.1$~kpc), random orbital energy and orbital angular momentum.  With these things held constant between test particles, the value of $\Phi_{eff}$ is not the same for different initial azimuthal coordinates, because $\Phi_{eff}=\Phi_{eff}(R,\phi)$.  Consequently, the value of Jacobi integral, and thus the value of $\Lambda_{nc,2}(t=0)$, for each test particle is different and therefore two test particles under these conditions might not both have the same trapped status. The $m=2$ spiral perturbation used in fig.~\ref{fig:phi0_vs_trapped} has $R_{CR}=9.5$~kpc and is otherwise the same prescription used in fig.~\ref{fig:CR8_circular}.  The initial conditions are modified from fig.~\ref{fig:CR8_circular} such that each test particle is launched from the apocentre of its epicycle with the value of the initial velocity being $(v_{ran,0,R},v_{ran,0,\phi})=(0,-10)$~km~s$^{-1}$.  The initial azimuthal coordinates in the three sets of panels in fig.~\ref{fig:phi0_vs_trapped} are $\phi_{1,0}=\lbrace\phi_{max},\phi_{min}/2,\phi_{min}\rbrace=\lbrace0,\pi/4,\pi/2\rbrace$ (from top to bottom).  In the left panels, we plot the contour for which $\Phi_{eff}=E_J$ (the ZVC) as a thin, dashed line, when a ZVC exists.  In the top and middle sets of panels, the capture criterion is met ($\Lambda_{nc,2}(t=0)=0.8$ and~$0.0$, respectively) and the test particle is in a trapped orbit.  In the bottom set of panels, $\Lambda_{nc,2}(t=0)=-1.1$  and the test particle is \textit{not} in a trapped orbit and instead circles the galactic center in the rotating frame. We again note that in all cases where the test particle is in a trapped orbit, the guiding centre radius (thick, dotted, black curve) is within the capture region, and in all cases where the test particle is not trapped it is not in the capture region.  The results in Fig.~\ref{fig:phi0_vs_trapped} demonstrate that the range in $R_L$ for which a star meets the capture criterion is dependent on azimuthal position, as one might expect from the radial thickness of the capture region as a function of azimuth.  This matches the prediction of the capture criterion (eqn.~\ref{eqn:trappedMeFlatEpiLz}) at $t=0$ printed in the panels on the right.

The value of the quantity $\Lambda_{nc,2}(t=0)$ in Figs.~\ref{fig:CR8_circular}-\ref{fig:phi0_vs_trapped} is much better at predicting (via eqn.~\ref{eqn:trappedMeFlatEpiLz}) whether or not a test particle with some finite value for $E_{ran}$ is in a trapped orbit than is $\Lambda_c$ (criteria for stars with $E_{ran}=0$), as is expected.  In the case that $E_{ran}=0$, $\Lambda_{nc,2}(t=0)=\Lambda_c$ and both are valid capture criteria.  Although the derivation of $\Lambda_{nc,2}$ assumes the epicyclic approximation, our tests (not all shown) suggest that the value of $\Lambda_{nc,2}(t)$ is a valid predictor of \lq\lq trapping" for stars on non-circular orbits where the unperturbed orbital trajectory would not be well described by the epicyclic approximation (e.g.~radial excursions in fig.~\ref{fig:CR8_randomvR} are on order of a scale length).  We therefore conclude that the value of $\Lambda_{nc,2}(t)$ (via eqn.~\ref{eqn:trappedMeFlatEpiLz}) is a robust indicator for whether or not a star is in a trapped orbit, while $\Lambda_c$ (via eqn.~\ref{eqn:trapped}) is only able to predict whether or not a star on an orbit with no random energy is in a trapped orbit.  We further explore our capture criteria in the event of scattering in the next subsection (\S\ref{sec:Scattering}).

In all the cases we tested numerically (not all of which are shown), a star is captured in a trapped orbit when its guiding centre radius (not the star itself) is within the capture region.  Further, a test particle that is not captured in a trapped orbit has its guiding centre radius outside the capture region.  In the limit that a star has zero random energy (i.e., $R=R_L$) it is in a trapped orbit when the star itself is positioned in the capture region.  There is no such requirement for the trajectory of test particles with some finite random orbital energy since the guiding center of the star is not equal to its coordinate position.  A trapped test particle may have a trajectory that physically leaves the capture region, while its guiding centre radius remains within the capture region (e.g.~fig.~\ref{fig:CR8_randomvphi}).  A test particle that is not captured in a trapped orbit may enter the capture region (e.g.~fig.~\ref{fig:CR8_randomvR}), while its guiding centre radius does not enter the capture region.  We, therefore, confirm our earlier prediction (\S\ref{sec:CaptureWithEpi}) that it is largely orbital angular momentum that determines whether or not a star is in a trapped orbit.

\subsection{Scattering}\label{sec:Scattering}

We defined scattering in \S\ref{sec:DerivationWRanMotions} as any process that changes both a star's orbital angular momentum and random orbital energy \citep[e.g.][]{SS53, BW67, Wielen77}.  This is in contrast to the oscillatory changes in the orbital angular momentum of a star in a trapped orbit, that are not accompanied by significant changes random orbital energy.  Up to this point we have not discussed what happens to a star that is in a trapped orbit when it is scattered.
  
A star is in a trapped orbit when the amplitude of the quantity $\Lambda_{nc}(t)$ (eqn.~\ref{eqn:Lambda_nc}) is less than unity (by eqn.~\ref{eqn:trappedMe}).  This quantity depends on both a star's orbital angular momentum and random orbital energy, and is not conserved.  A star on an orbit that initially meets the capture criterion (eqn.~\ref{eqn:trappedMeFlatEpi}) will continue to do so unless it is scattered such that $|\Lambda_{nc,\beta}(t)|$ becomes greater than unity, at which time the star will no longer be in a trapped orbit.

Our orbital integrator uses a smooth underlying potential, with a superposed spiral perturbation that has constant amplitude for the duration of the orbital integration.  This is appropriate for testing the validity of eqn.~\ref{eqn:trappedMeFlatEpiLz}, but such a potential is also somewhat unrealistic.  It is well established that small-scale fluctuations in the galactic potential (such as GMCs) \citep{SS53,Wielen77,Lacey84} as well as short lived, physically extended fluctuations (such as transient spiral arms) \citep*{BW67,CS85,DWT04} lead to scattering of stellar orbits, altering random orbital energies and orbital angular momenta.  It is therefore important to understand that stars we find to remain in trapped orbits indefinitely may not be representative of stellar behaviour in a lumpy (i.e.~small scale-length fluctuations in the potential), time-dependent underlying potential, where we expect that many of these stars would be scattered out of their trapped orbits.

Despite the rather smooth potential we assume, we are able to observe the effects of scattering on the orbital trajectory of a test particle and its value of $\Lambda_{nc,2}(t)$.  We observe two processes by which stars are scattered (illustrated below).

First, a star in a non-axisymmetric potential is scattered whenever it is not in instantaneous corotation with the perturbing pattern \citep[by eqn.~\ref{eqn:SB02JR}, and eqn.~4 in][]{SB02}.  There is a range of values for the guiding centre radius, $R_L(t)$ (and therefore a range in $\Omega_g$), that satisfies the capture criterion (eqn.~\ref{eqn:trappedMeFlatEpiLz}), not simply the $0^{th}$ order assumption that $R_L=R_{CR}$.  Indeed, a star in a trapped orbit has an oscillating value for $R_L(t)$ (i.e. $\Omega_g(t))$.  Therefore, other than when a star's guiding centre radius equals the radius of corotation - which happens twice during its trapped orbit - the star will experience changes in its orbital angular momentum that are accompanied by changes in random orbital energy (see eqn.~\ref{eqn:SB02JR}).  This type of scattering becomes particularly important when the spiral wave number is large (i.e.~the radial spacing between spiral arms is small) and a star particle that is in a trapped orbit has a close approach to a spiral arm away from corotation.  Through such interactions away from corotation, a star particle that is initially captured in trapped orbits may begin circling the galactic centre in the rotating frame (i.e.~no longer in a trapped orbit).\footnote{The first order change to a phase space \textit{distribution function} tends to zero away from spiral resonances \citep{CS85}.}  We defer a discussion of the time-scale for this type of scattering out of a trapped orbit to Paper~III.

Second, changes to a star's random orbital energy ($E_{ran}(t)$) are most dramatic at the Lindblad resonances (the radii at which $\kappa=\pm m(\Omega_p-\Omega_g)$), where the star passes (or is passed by) the perturbation at the star's epicyclic frequency.  One would expect enhanced scattering should a star in a trapped orbit cross a Lindblad resonance.  The capture region can overlap with a Lindblad resonance when either, (1) the radial range of the capture region is large (e.g.~the amplitude of the spiral potential is high), or (2) the Lindblad resonances are close to corotation (e.g.~for values of $R_{CR}$ that are close to the galactic centre).   We observe erratic changes in the motion when the guiding centre radius ($R_L(t)$) of a star in a trapped orbit encounters a Lindblad resonance (including the ultraharmonic Lindblad resonances, where $\kappa=\pm 2m(\Omega_p-\Omega_\phi)$).  \cite{Chirikov79} predicted that chaotic behaviour would emerge when an object enters a region of resonant overlap in phase space (i.e.~the region in a surface of section where two resonances occupy the same space).  Irregular orbital motions (sometimes called \lq\lq wild" or ergodic) have been observed to arise in simulations when stars pass through regions of resonant overlap in phase space \citep[eg.][]{Martinet74,Athanassoula83,Pichardo03}; where \cite{Pichardo03} used a Lyapunov exponent analysis to identify these motions as chaotic.

In the present case, we are observing resonant overlap in coordinate space.  The boundary between orbits that circle the galactic center in the rotating frame and orbits that are trapped is called the \lq\lq separatix" in a surface of section diagram.  In the case of zero random energy, the separatix can be projected onto coordinate space as a contour that encloses the capture region.  As per the discussion in \S\ref{sec:NumericalTestingCaptureCriteria}, stars with some finite random orbital energy that are in trapped orbits have guiding centre radii that are inside the capture region.  We do not quantitatively prove that we are observing the emergence of chaotic behaviour at resonant overlap, however, our tests are consistent with this theory as there is a sudden increase in random energy and change in orbital trajectory when the guiding center radius of a trapped star encounters a Lindblad resonance.

\begin{figure}
\begin{center}
\includegraphics[scale=0.5]{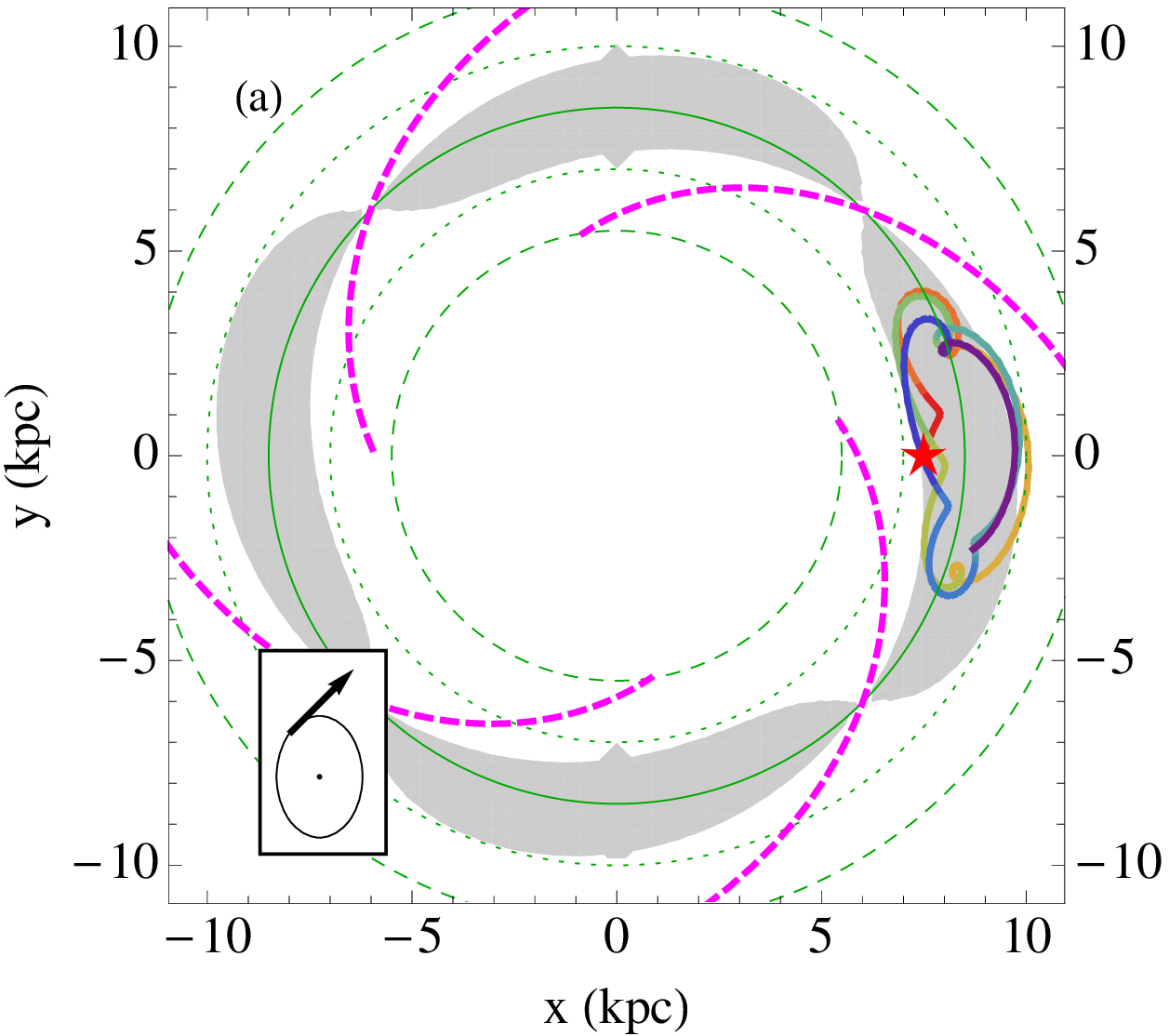}
\includegraphics[scale=0.5]{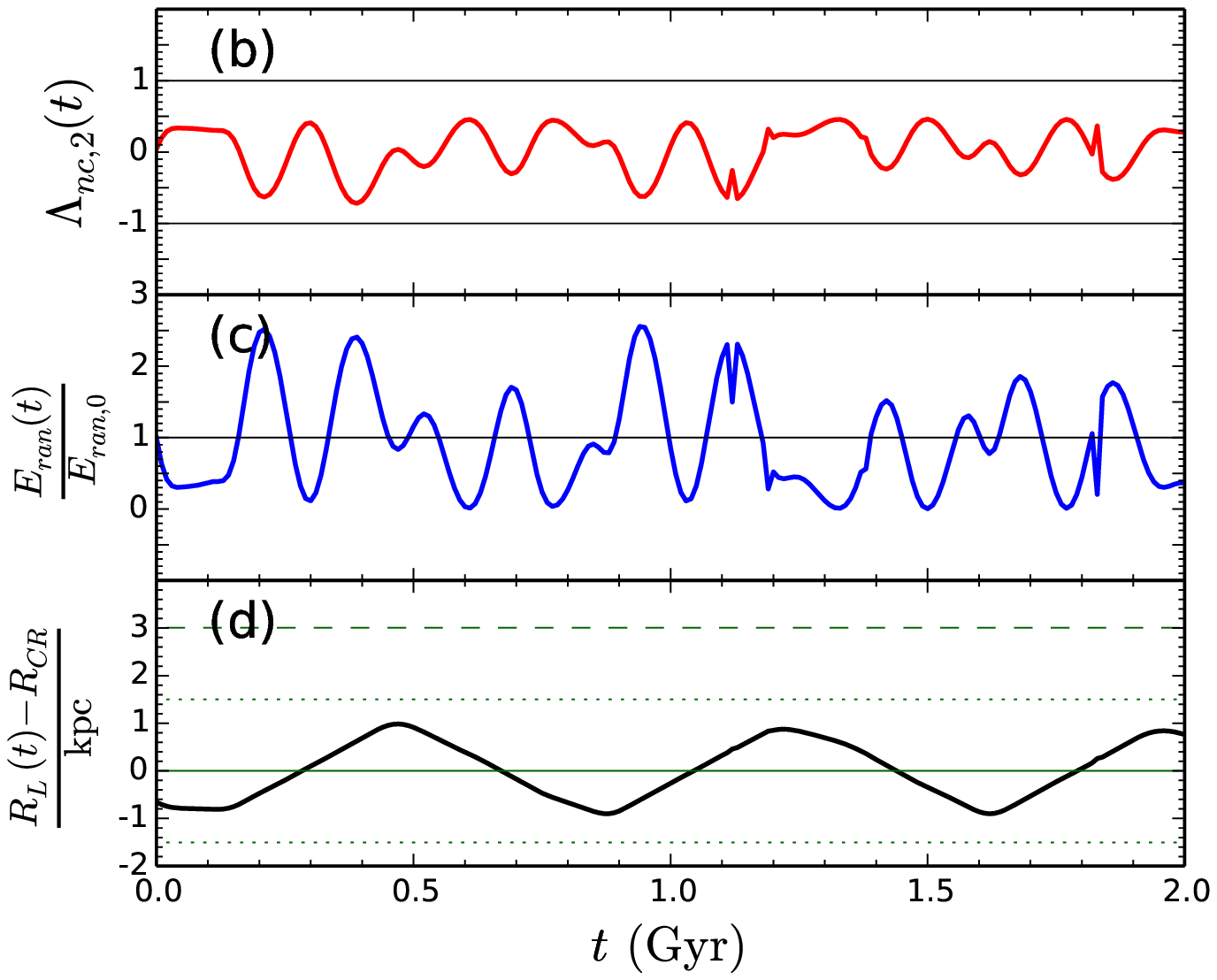}
\caption{Orbital properties of a star particle launched at $1$~kpc inside corotation with $(v_{ran,R,0},v_{ran,\phi,0})=(10,10)$~km~s$^{-1}$ ($R_{L,0}=7.8$~kpc) in a potential modified from Fig.~\ref{fig:CR8_circular} so that $R_{CR}~=~8.5$~kpc.  Panel~(a) is the same as in Fig.~\ref{fig:CR8_circular}.  Panel~(b) shows $\Lambda_{nc,2}(t)$ as a solid red line. The horizontal, grey lines at $1$ and $-1$ are the upper and lower limits for $\Lambda_{nc,2}(t)$ for the star to be in a trapped orbit.  Note that the star is in a trapped orbit for the entire integration and therefore $|\Lambda_{nc,2}(t)|<1$ at all times.  Panel~(c) shows the ratio ($E_{ran}(t)/E_{ran,0}$) as a solid, blue line.  Panel~(d) shows $(R_L(t)-R_{CR})$~[kpc] as a solid, black line.  In panels~(a) and (d), the dashed and dotted dark-green curves show the Lindblad resonances (where $\kappa=\pm m(\Omega_p-\Omega)$) and the ultra-harmonic Lindblad resonances (where $\kappa=\pm 2m(\Omega_p-\Omega)$), respectively.  This star is in a trapped orbit.}
\label{fig:Scatter1}
\end{center}
\end{figure}

\begin{figure}
\begin{center}
\includegraphics[scale=0.5]{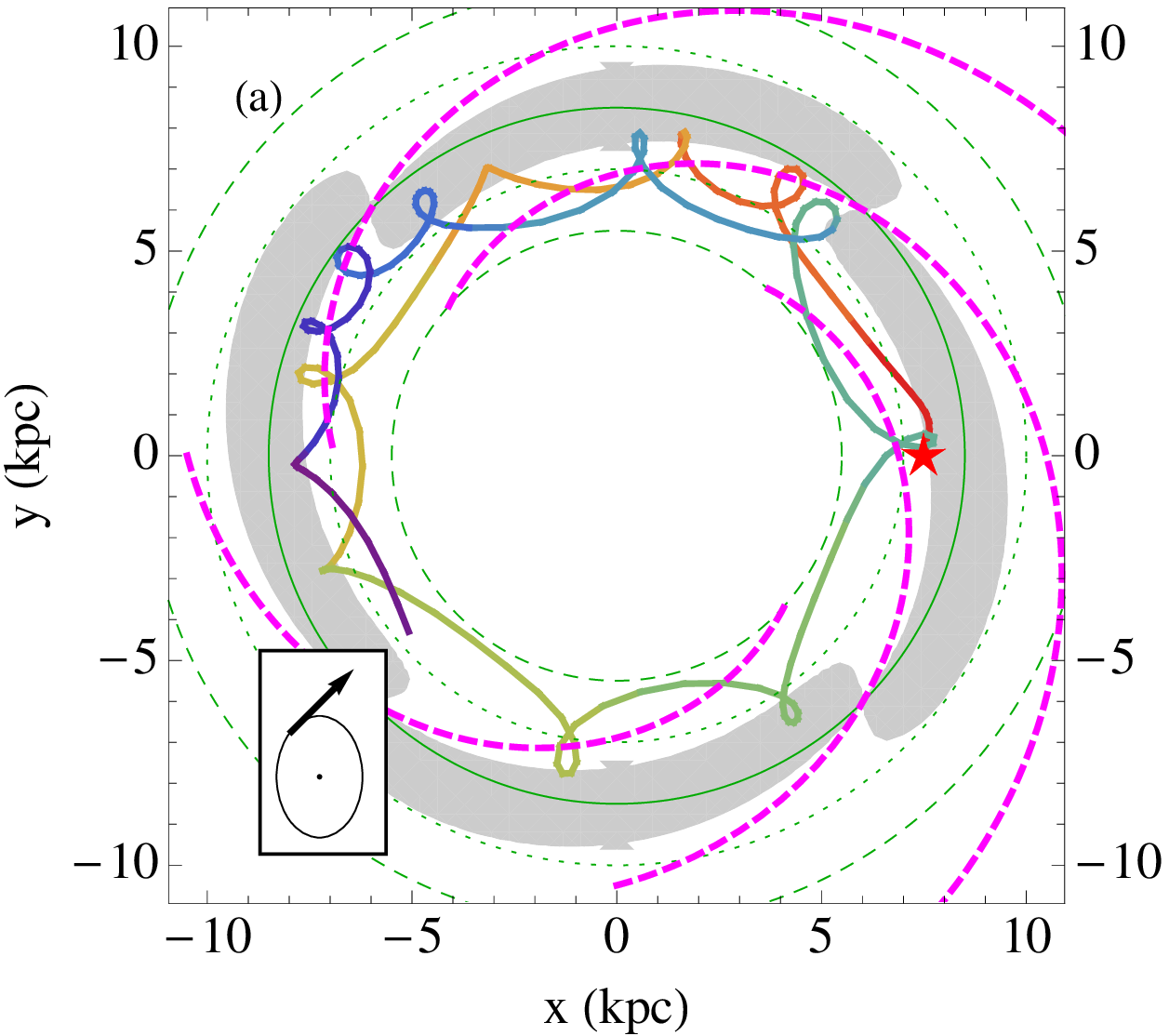}
\includegraphics[scale=0.5]{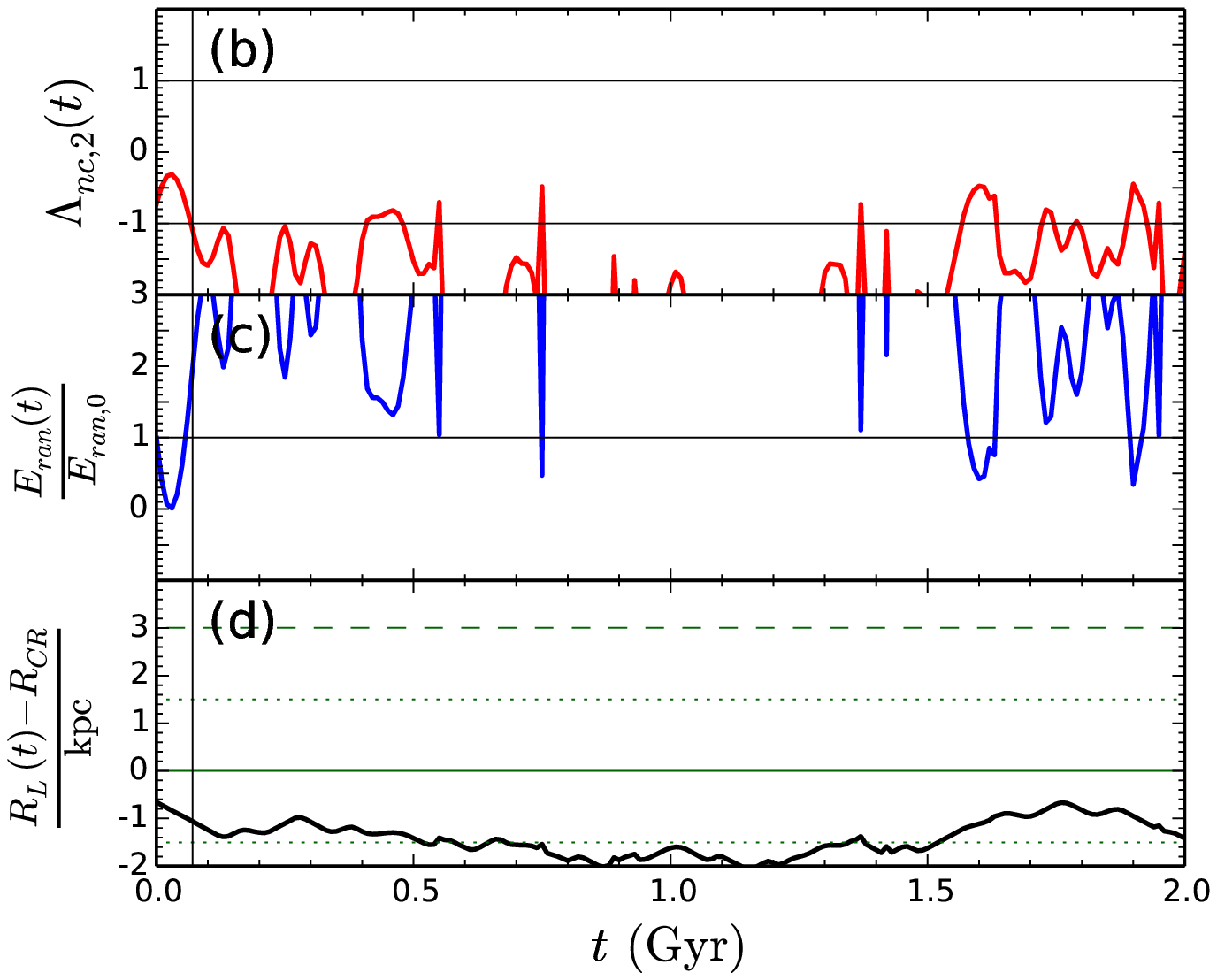}
\caption{Panels as in Fig.~\ref{fig:Scatter1}.  Orbital properties of a star particle with the potential modified from that in Fig.~\ref{fig:Scatter1} so that $R_{CR}=8$~kpc.  The radial coordinate of the star at $t=0$ is $1$~kpc inside corotation and its guiding centre radius is $R_{L,0}=7.3$~kpc.  This star is initially in a trapped orbit, but as $E_{ran}(t)$ increases, the star scatters early in the integration interval such that $\Lambda_{nc,2}(t) < -1$ marked with vertical line and is no longer in a trapped orbit likely corresponding to the star's close approach to the spiral arm away from corotation.}
\label{fig:Scatter2}
\end{center}
\end{figure}

\begin{figure}
\begin{center}
\includegraphics[scale=0.5]{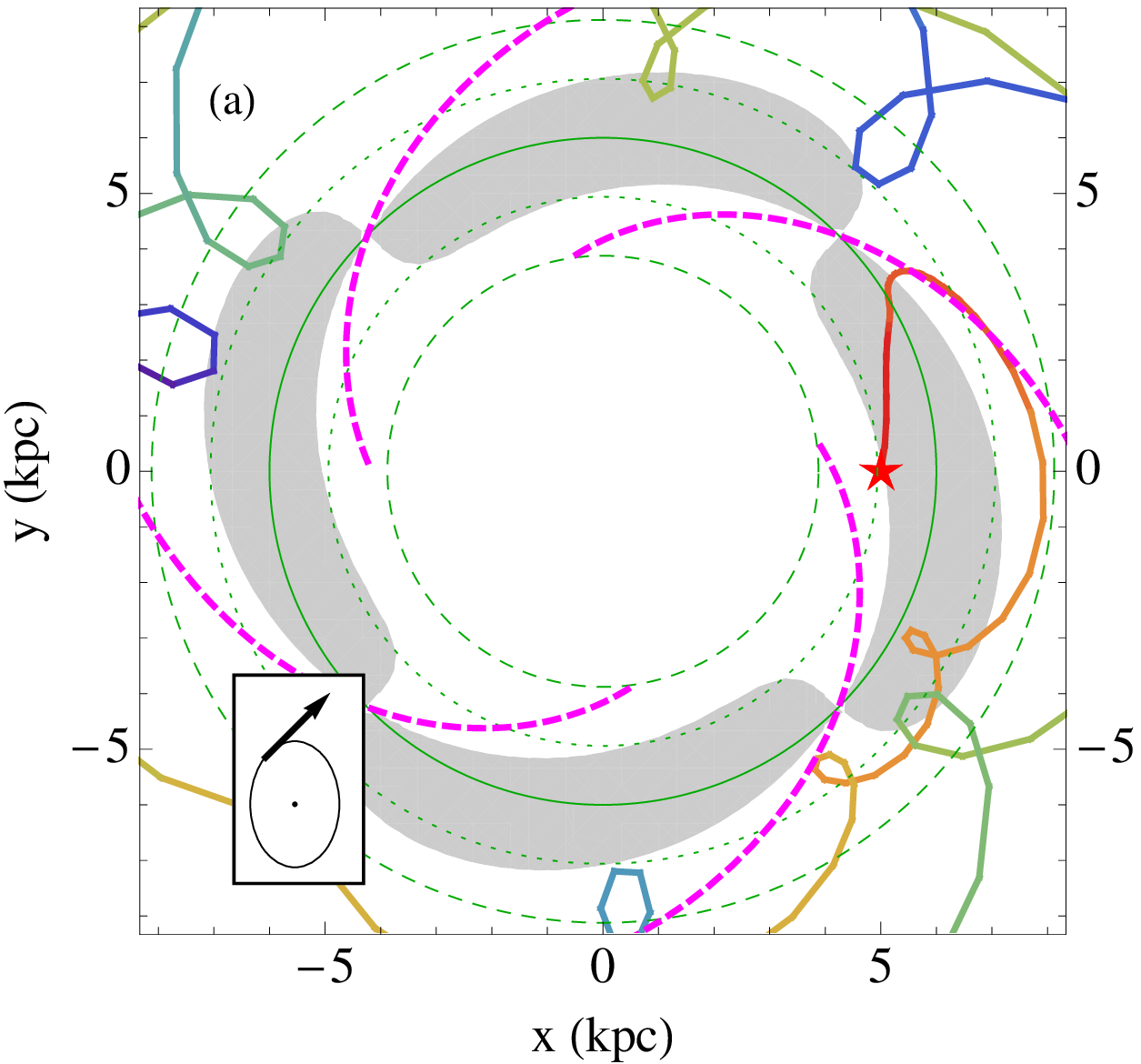}
\includegraphics[scale=0.5]{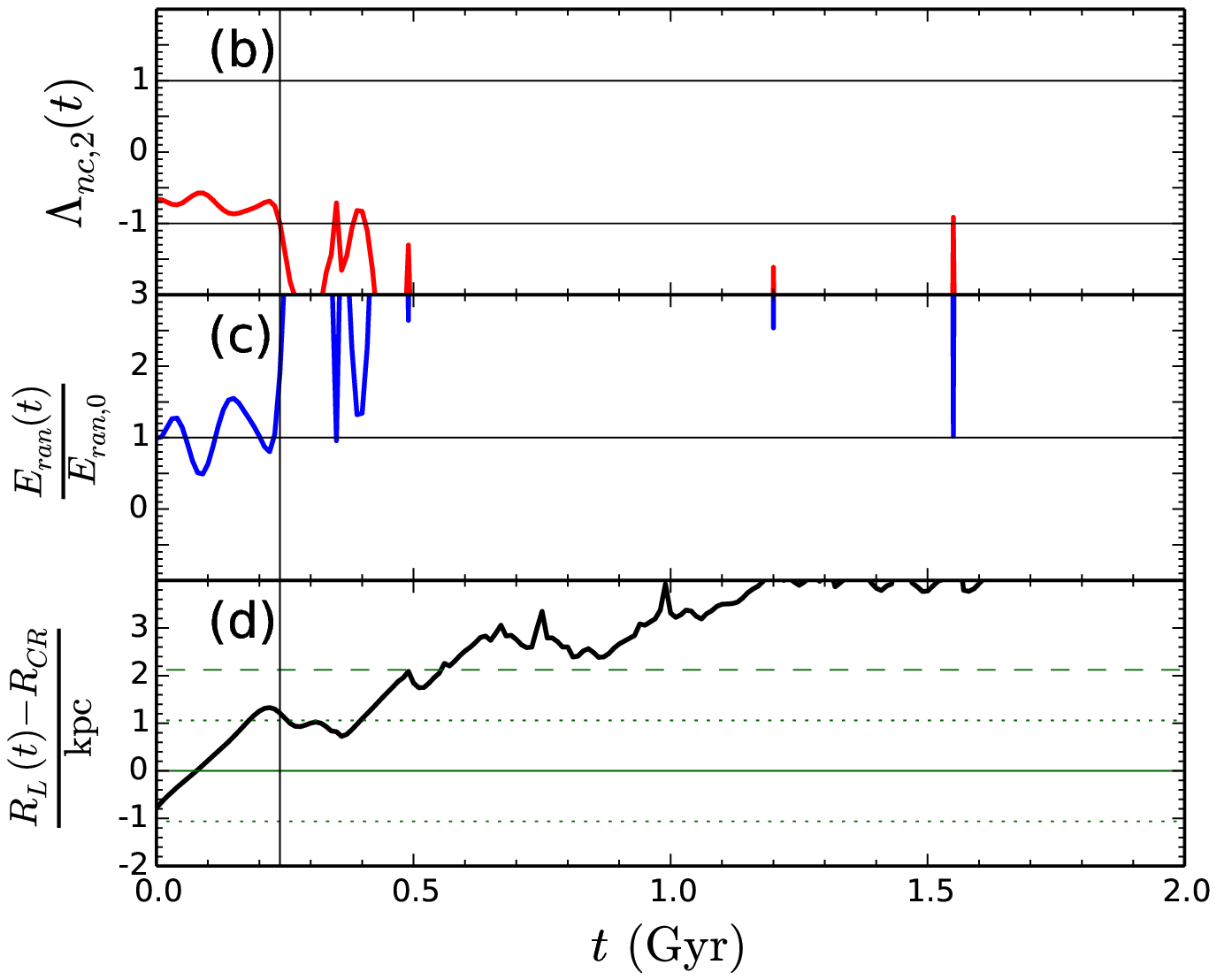}
\caption{Panels as in Fig.~\ref{fig:Scatter1}.  Orbital properties of a star particle with initial conditions modified from those in Fig.~\ref{fig:Scatter1} so that $R_{CR}=6$~kpc. The test particle is launched $1$~kpc inside corotation and its guiding centre radius is at $R_{L,0}=5.2$~kpc.  This star initially meets the capture criterion, but rapidly scatters when the star's guiding centre radius crosses the first ultra-harmonic outer Lindblad resonance.  A vertical line indicates the time when the star is no longer captured in a trapped orbit.}
\label{fig:Scatter3}
\end{center}
\end{figure}

Figures~\ref{fig:Scatter1},~\ref{fig:Scatter2},~\&~\ref{fig:Scatter3} show the orbits of test particles for three sets of initial conditions.  Each test particle initially meets the capture criterion (eqn.~\ref{eqn:trappedMeFlatEpiLz}) and is launched with initial random velocity $(v_{ran,0,R},v_{ran,0,\phi})=(10,10)$~km~s$^{-1}$ at an initial radius $1$~kpc inside the radius of corotation in a spiral potential with $m=4$, $\epsilon_\Sigma=0.3$, and $\theta=25^\circ$ for $R_{CR}=\left\lbrace 8.5,8.0,6.0\right\rbrace$~kpc respectively.  For each set of initial conditions, panel (a) shows the orbital path in the rotating frame, as in Fig.~\ref{fig:CR8_circular}.  Panel (b) shows the time evolution of $\Lambda_{nc,2}(t)$ (solid, red). The horizontal black lines at $\Lambda_{nc,2}(t)=1$ and $-1$ are the upper and lower limits for trapped orbits.  Panel (c) shows the random orbital energy normalised by its initial value ($E_{ran}(t)/E_{ran,0}$) (solid, blue).  The horizontal black line at $E_{ran}(t)/E_{ran,0}=1$ indicates the initial value.  Panel (d) shows $(R_L(t)-R_{CR})$ (solid, black).  In both panes (a) and (d), dashed, dark-green curves show the inner and outer Lindblad resonances (ILR/ORL), and the dotted dark-green lines show the first ultra-harmonic Lindblad resonances (these are the 2$m$:1 resonances for an $m$-armed perturbation).  Figure~\ref{fig:Scatter1} shows a star particle in a trapped orbit that is not significantly scattered.  Figures~\ref{fig:Scatter2}~\&~~\ref{fig:Scatter3} show star particles that are initially in a trapped orbit but are scattered such that they no longer meet the capture criterion.  The vertical black line in panels (b)-(d) marks the time when $|\Lambda_{nc,2}(t)|>1$, and thus the star particles are no longer in trapped orbits.  The star particle in figure~\ref{fig:Scatter2} is scattered out of a trapped orbit when the star approaches a spiral arm away from corotation (instantaneous $R_L\neq R_{CR}$).  As can be seen in panel~(d), the vertical black line does not correspond to a time when the guiding centre radius crosses a Lindblad radius, but rather when the star has an increase in random orbital energy as it approaches the peak density of the spiral perturbation away from corotation.  Figure~\ref{fig:Scatter3} shows a star particle that initially meets the capture criterion but is scattered out of a trapped orbit (marked by the vertical black line in the right-hand panels).  In panel~(d), it is clear that this corresponds to the time when the guiding center radius ($R_L(t)$) crosses the ultra-harmonic OLR (dotted, horizontal line).  These plots illustrate that a star with some finite random orbital energy, which is initially in a trapped orbit, can be scattered leading to $|\Lambda_{nc,2}(t)|>1$.  Although we do not show an example here, it is possible that a star that is not initially in a trapped orbit could be captured in a trapped orbit if it is scattered in such a way that eqn.~\ref{eqn:trappedMeFlatEpiLz} is satisfied.

A capture region with a larger area, which grows with spiral strength (see discussion in \S\ref{sec:CaptureRegion}), will be able to support trapped orbits for stars with guiding center radii farther from corotation.  The radial range of the capture region is important since a star's random energy slowly increases with radial distance from corotation and since a broad capture region is more likely to overlap with a Lindblad resonance.  Therefore the amplitude of our chosen spiral perturbation to the potential (eqn.~\ref{eqn:SpiralAmplitude}) is greater for spirals that have a high fractional amplitude in surface density ($\epsilon_\Sigma$), in regions of the disk that have a small wave number ($k(R)= m \cot(\theta) R^{-1}$), and as $|R-R_d|\rightarrow 0$ in a disk with an exponential surface density since $\Phi_s \propto R \Sigma(R)$.

In a lumpy underlying potential (unlike the one we use), a star that is in a trapped orbit will also have gravitational interactions with fluctuations in the potential other than the spiral pattern.  These interactions will cause the star to have changes in its orbital angular momentum and random energy \citep[see][and references therein]{Sellwood13} that may cause the star to no longer meet the capture criterion.

Finally, resonant overlap can also occur when there are two (or more) perturbations to the underlying axisymmetric potential with different pattern speeds.  \cite{MF10} simulated \citep[further explored by][]{Minchev11} a disk with both a bar and spiral perturbation.  They found that when resonances of the two patterns overlapped, the changes to the orbital angular momentum of disk stars were greater than the sum of the separate changes in orbital angular momentum from the individual non-axisymmetric patterns.  In the context of the present paper, we would expect irregular orbits to emerge in the case that the OLR of a central bar pattern overlaps the capture region from the spiral pattern.

\section{Capture for a Non-Steady Spiral Pattern}\label{sec:Discussion}

In this paper, we have considered the case of a spiral perturbation with a pattern speed that is independent of radius and time.  However, N-body simulations of disk galaxies frequently exhibit spiral arms that appear to have pattern speeds that are radially dependent (e.g.~\cite*{GKC12}, \cite{BSW13}, \cite{R-F13}; but see \cite{SC14} for an alternative interpretation) or that evolve in time \citep[e.g.][]{Roskar12}.  We now explore how radial migration would differ from the analysis in this paper should either scenario be the case.

In Appx.~\ref{sec:AppxA}, we explore the difference between two different capture criteria for stars with zero random energy.  Eqns.~\ref{eqn:A1Capture} and \ref{eqn:A2CaptureReduced} relate the radial range for a star with zero random orbital energy to be in a trapped orbit (and therefore the size of the capture region) to the shape of the rotation curve given the spiral has a constant pattern speed.  The nature of this relationship is most transparent in eqn.~\ref{eqn:BTcapture}, but can be obtained from eqn.~\ref{eqn:Lambda_ncbeta} or eqn.~\ref{eqn:A2CaptureLong} with appropriate choice of the functional form of $\dot{x}_\phi$.  Eqn.~\ref{eqn:BTcapture} can be rearranged to solve for the radial range of the capture region (ie. the maximum value for $|R(t)-R_{CR}|$) in the effective potential for a given shape of the rotation curve.  For the same amplitude of the perturbation ($|\Phi_b|$) and pattern speed ($\Omega_p$), a central point potential has the smallest radial range for the capture region, while stars in solid body rotation are in corotation at all $R$ and therefore are captured at all radii.  The corollary is that the radial range of the capture region is determined by the rate of divergence with distance from corotation between the possibly radially dependent pattern speed ($\Omega_p(R)$) and the circular orbital frequency ($\Omega_c(R)$).  Should the functional form of the pattern speed approach that of the circular orbital frequency, the radial range for capture would include the entire disk.  In support of this interpretation, \cite{GKC12} found that in a 3D N-body simulation of a disk with transient spiral arms with $\Omega_p(R)\approx\Omega(R)$, stars from a wide range of initial radii migrated along the spiral arms, sometimes having $10-20\%$ changes in angular momentum.

In the case of steady spiral patterns, very efficient radial migration requires multiple transient spiral patterns, with a random distribution of pattern speeds, over the lifetime of the disk.  In such a scenario, a star will migrate radially as a random walk, where the size of any step is determined by the amplitudes of the perturbations \citep[see][]{SB02} and the number of steps is related to the duty cycle of transient spirals.  However, should a spiral pattern speed be time-dependent, a small amplitude spiral could cause a star to migrate over a large radial distance.  To understand this, consider a disk with a flat rotation curve and a spiral pattern speed that decreases in time, thus causing the radius of corotation to increase with time and the location of the capture region to move to ever larger radii.  Should the guiding center radius of a trapped star be increasing at the same rate as the radius of corotation, the star could continuously migrate outward for the lifetime of the spiral pattern.  In the same scenario, a star which had a different initial phase in its trapped orbit such that its guiding center radius is decreasing might only be trapped briefly as the capture region moves outward.  Such time dependent pattern speeds could cause the ensemble of disk stars to migrate longer distances preferentially outward (or inward in the case of a pattern speed that increases with time), and even a low amplitude spiral could lead to large radial excursions for such stars.  There are examples of time-dependent spiral structure in simulations of spiral galaxies \citep[e.g.][]{Roskar12,SC14}, but it is unclear whether or not these patterns host stars that migrate large radial distances.

\section{Conclusions}\label{sec:Conclusions}
This is the first of a series of papers addressing the physical parameters important to the efficiency of radial migration in disk galaxies (i.e.~changes to stellar angular momentum around corotation without associated kinematic heating).  We focus on the conditions necessary for a star to be in a trapped orbit near corotation.  A trapped orbit (defined in \S\ref{sec:Capture}), caused by changes in orbital angular momentum from gravitational torques by spiral arms, describes the motion of a star's guiding centre radius as it oscillates across and back through the radius of corotation of a spiral perturbation.  Should the spiral be transient, a disk star in a trapped orbit could migrate radially, i.e.~have a long-lived change in its mean orbital radius reflecting the change in angular momentum. 

We derive the \lq\lq capture criterion" that determines whether or not a star with some finite radial action (which may be related to orbital random energy by eqn.~\ref{eqn:SB02JR}) in a 2D disk is in a trapped orbit (eqns.~\ref{eqn:Lambda_nc}~\&~\ref{eqn:trappedMe}). This is in contrast to the capture criterion for a 2D disk in the literature \citep{Contopoulos78}, which assumes zero random orbital energy.  We further derive a general expression for the capture criterion in terms of a star's random orbital energy ($E_{ran}(t)$) and orbital angular momentum ($L_z(t)$) in a disk where the underlying potential leads to a given rotation curve (eqn.~\ref{eqn:trappedMeFlatEpi}).  In a disk with a flat rotation curve the capture criterion ($\Lambda_{nc,2}$) is well described by eqn.~\ref{eqn:trappedMeFlatEpiLz}.

We find that orbital angular momentum is the most important factor in determining whether or not a star in the disk is in a trapped orbit, while random orbital energy is less influential.  We use the capture criterion to derive an expression for the region, called the \lq\lq capture region" (\S\ref{sec:CaptureRegion}), within which a star with zero random orbital energy must be located in order to be captured in a trapped orbit.  We propose that whether or not a star is in a trapped orbit is closely approximated by whether or not its guiding centre radius ($R_L(t)$) is within the capture region.  Radial excursions from random orbital energy may cause a star that is not in a trapped orbit to enter the capture region, or a star that is in a trapped orbit to leave the capture region, but the star's status as captured or not in a trapped orbit is not influenced by these excursions.

A star that is in a trapped orbit will remain in a trapped orbit indefinitely, unless the star is scattered.  We define scattering as any event that causes a star to experience a change in it guiding centre radius that is associated with a change in random orbital energy.  Thus, a scattering event can cause a star that is in a trapped orbit to no longer meet the capture criterion.  In an inhomogeneous potential with multiple, small-scale length perturbations, as opposed to the potential we assume for the current study, it is more likely that a star in a trapped orbit will be scattered.  We find that when the guiding centre radius of a star in a trapped orbit approaches a Lindblad resonance, the star is rapidly scattered out of a trapped orbit.  Since the parameter $\Lambda_{nc}(t)$ is a time-dependent quantity, it is important to realize that a star which initially meets the capture criterion may not remain in a trapped orbit for long enough to migrate radially.  

\section*{Acknowledgments}
It is a pleasure to thank Jerry Sellwood, Barbara Pichardo and Jonathan Bird for helpful discussions during the course of this work.  We thank the anonymous referee for comments which strengthened the discussion in this paper.  This material is based upon work supported by the National Science Foundation Graduate Research Fellowship under Grant No.~DGE-1232825 and National Science Foundation Grants AST-0908326 and OIA-1124403.

\bibliographystyle{mn2e} 
\bibliography{mybibliography.bib}

\begin{thebibliography}{}

\bibitem[\protect\citeauthoryear{{Athanassoula}, {Bienayme}, {Martinet} \&
  {Pfenniger}}{{Athanassoula} et~al.}{1983}]{Athanassoula83}
{Athanassoula} E.,  {Bienayme} O.,  {Martinet} L.,    {Pfenniger} D.,  1983,
  \aap, 127, 349

\bibitem[\protect\citeauthoryear{{Baba}, {Saitoh} \& {Wada}}{{Baba}
  et~al.}{2013}]{BSW13}
{Baba} J.,  {Saitoh} T.~R.,    {Wada} K.,  2013, \apj, 763, 46

\bibitem[\protect\citeauthoryear{{Barbanis}}{{Barbanis}}{1976}]{Barbanis76}
{Barbanis} B.,  1976, Celestial Mechanics, 14, 201

\bibitem[\protect\citeauthoryear{{Barbanis} \& {Woltjer}}{{Barbanis} \&
  {Woltjer}}{1967}]{BW67}
{Barbanis} B.,  {Woltjer} L.,  1967, \apj, 150, 461

\bibitem[\protect\citeauthoryear{{Binney} \& {Tremaine}}{{Binney} \&
  {Tremaine}}{1987}]{BT87}
{Binney} J.,  {Tremaine} S.,  1987, {Galactic dynamics}.
Princeton University Press

\bibitem[\protect\citeauthoryear{{Binney} \& {Tremaine}}{{Binney} \&
  {Tremaine}}{2008}]{BT08}
{Binney} J.,  {Tremaine} S.,  2008, {Galactic Dynamics: Second Edition}.
Princeton University Press

\bibitem[\protect\citeauthoryear{{Bird}, {Kazantzidis} \& {Weinberg}}{{Bird}
  et~al.}{2012}]{BKW12}
{Bird} J.~C.,  {Kazantzidis} S.,    {Weinberg} D.~H.,  2012, \mnras, 420, 913

\bibitem[\protect\citeauthoryear{{Brown}}{{Brown}}{1911}]{Brown11}
{Brown} E.~W.,  1911, \mnras, 71, 438

\bibitem[\protect\citeauthoryear{{Carlberg} \& {Sellwood}}{{Carlberg} \&
  {Sellwood}}{1985}]{CS85}
{Carlberg} R.~G.,  {Sellwood} J.~A.,  1985, \apj, 292, 79

\bibitem[\protect\citeauthoryear{{Chirikov}}{{Chirikov}}{1979}]{Chirikov79}
{Chirikov} B.~V.,  1979, Phys.Rep, 52, 263

\bibitem[\protect\citeauthoryear{{Contopoulos}}{{Contopoulos}}{1973}]{Contopoulos73}
{Contopoulos} G.,  1973, \apj, 181, 657

\bibitem[\protect\citeauthoryear{{Contopoulos}}{{Contopoulos}}{1975}]{Contopoulos75}
{Contopoulos} G.,  1975, \apj, 201, 566

\bibitem[\protect\citeauthoryear{{Contopoulos}}{{Contopoulos}}{1978}]{Contopoulos78}
{Contopoulos} G.,  1978, \aap, 64, 323

\bibitem[\protect\citeauthoryear{{De Simone}, {Wu} \& {Tremaine}}{{De Simone}
  et~al.}{2004}]{DWT04}
{De Simone} R.,  {Wu} X.,    {Tremaine} S.,  2004, \mnras, 350, 627

\bibitem[\protect\citeauthoryear{{Debattista}, {Mayer}, {Carollo}, {Moore},
  {Wadsley} \& {Quinn}}{{Debattista} et~al.}{2006}]{Debattista06}
{Debattista} V.~P.,  {Mayer} L.,  {Carollo} C.~M.,  {Moore} B.,  {Wadsley} J.,
    {Quinn} T.,  2006, \apj, 645, 209

\bibitem[\protect\citeauthoryear{{Dehnen}}{{Dehnen}}{1999}]{Dehnen99}
{Dehnen} W.,  1999, \aj, 118, 1190

\bibitem[\protect\citeauthoryear{{Fran\c{c}ois} \& {Matteucci}}{{Fran\c{c}ois}
  \& {Matteucci}}{1993}]{FM93}
{Fran\c{c}ois} P.,  {Matteucci} F.,  1993, A\& A, 280, 136

\bibitem[\protect\citeauthoryear{{Goldreich} \& {Tremaine}}{{Goldreich} \&
  {Tremaine}}{1982}]{GT82}
{Goldreich} P.,  {Tremaine} S.,  1982, \araa, 20, 249

\bibitem[\protect\citeauthoryear{{Grand}, {Kawata} \& {Cropper}}{{Grand}
  et~al.}{2012}]{GKC12}
{Grand} R.~J.~J.,  {Kawata} D.,    {Cropper} M.,  2012, \mnras, 421, 1529

\bibitem[\protect\citeauthoryear{{Kormendy}}{{Kormendy}}{2013}]{Kormendy13}
{Kormendy} J.,  2013, {Secular Evolution in Disk Galaxies}.
Cambridge University Press

\bibitem[\protect\citeauthoryear{{Lacey}}{{Lacey}}{1984}]{Lacey84}
{Lacey} C.~G.,  1984, \mnras, 208, 687

\bibitem[\protect\citeauthoryear{{Lin} \& {Shu}}{{Lin} \& {Shu}}{1964}]{LS64}
{Lin} C.~C.,  {Shu} F.~H.,  1964, \apj, 140, 646

\bibitem[\protect\citeauthoryear{{Lin}, {Yuan} \& {Shu}}{{Lin}
  et~al.}{1969}]{LYS69}
{Lin} C.~C.,  {Yuan} C.,    {Shu} F.~H.,  1969, \apj, 155, 721

\bibitem[\protect\citeauthoryear{{Loebman}, {Ro{\v s}kar}, {Debattista},
  {Ivezi{\'c}}, {Quinn} \& {Wadsley}}{{Loebman} et~al.}{2011}]{Loebman11}
{Loebman} S.~R.,  {Ro{\v s}kar} R.,  {Debattista} V.~P.,  {Ivezi{\'c}} {\v Z}.,
   {Quinn} T.~R.,    {Wadsley} J.,  2011, \apj, 737, 8

\bibitem[\protect\citeauthoryear{{Ma}}{{Ma}}{2002}]{Ma02}
{Ma} J.,  2002, \aap, 388, 389

\bibitem[\protect\citeauthoryear{{Martinet}}{{Martinet}}{1974}]{Martinet74}
{Martinet} L.,  1974, \aap, 32, 329

\bibitem[\protect\citeauthoryear{{Minchev} \& {Famaey}}{{Minchev} \&
  {Famaey}}{2010}]{MF10}
{Minchev} I.,  {Famaey} B.,  2010, ApJ, 722, 112

\bibitem[\protect\citeauthoryear{{Minchev}, {Famaey}, {Combes}, {Di Matteo},
  {Mouhcine} \& {Wozniak}}{{Minchev} et~al.}{2011}]{Minchev11}
{Minchev} I.,  {Famaey} B.,  {Combes} F.,  {Di Matteo} P.,  {Mouhcine} M.,
  {Wozniak} H.,  2011, \aap, 527, A147+

\bibitem[\protect\citeauthoryear{{Minchev}, {Famaey}, {Quillen}, {Dehnen},
  {Martig} \& {Siebert}}{{Minchev} et~al.}{2012}]{Minchev12b}
{Minchev} I.,  {Famaey} B.,  {Quillen} A.~C.,  {Dehnen} W.,  {Martig} M.,
  {Siebert} A.,  2012, \aap, 548, A127

\bibitem[\protect\citeauthoryear{{Papayannopoulos}}{{Papayannopoulos}}{1979a}]{Papayannopoulos79a}
{Papayannopoulos} T.,  1979a, \aap, 77, 75

\bibitem[\protect\citeauthoryear{{Papayannopoulos}}{{Papayannopoulos}}{1979b}]{Papayannopoulos79b}
{Papayannopoulos} T.,  1979b, \aap, 79, 197

\bibitem[\protect\citeauthoryear{{Pichardo}, {Martos}, {Moreno} \&
  {Espresate}}{{Pichardo} et~al.}{2003}]{Pichardo03}
{Pichardo} B.,  {Martos} M.,  {Moreno} E.,    {Espresate} J.,  2003, \apj, 582,
  230

\bibitem[\protect\citeauthoryear{{Rix} \& {Zaritsky}}{{Rix} \&
  {Zaritsky}}{1995}]{RZ95}
{Rix} H.-W.,  {Zaritsky} D.,  1995, \apj, 447, 82

\bibitem[\protect\citeauthoryear{{Roca-F{\`a}brega}, {Valenzuela}, {Figueras},
  {Romero-G{\'o}mez}, {Vel{\'a}zquez}, {Antoja} \&
  {Pichardo}}{{Roca-F{\`a}brega} et~al.}{2013}]{R-F13}
{Roca-F{\`a}brega} S.,  {Valenzuela} O.,  {Figueras} F.,  {Romero-G{\'o}mez}
  M.,  {Vel{\'a}zquez} H.,  {Antoja} T.,    {Pichardo} B.,  2013, \mnras, 432,
  2878

\bibitem[\protect\citeauthoryear{{Ro{\v s}kar}, {Debattista}, {Quinn} \&
  {Wadsley}}{{Ro{\v s}kar} et~al.}{2012}]{Roskar12}
{Ro{\v s}kar} R.,  {Debattista} V.~P.,  {Quinn} T.~R.,    {Wadsley} J.,  2012,
  \mnras, 426, 2089

\bibitem[\protect\citeauthoryear{{Ro{\v s}kar}, {Debattista}, {Stinson},
  {Quinn}, {Kaufmann} \& {Wadsley}}{{Ro{\v s}kar} et~al.}{2008}]{Roskar08a}
{Ro{\v s}kar} R.,  {Debattista} V.~P.,  {Stinson} G.~S.,  {Quinn} T.~R.,
  {Kaufmann} T.,    {Wadsley} J.,  2008, ApJL, 675, L65

\bibitem[\protect\citeauthoryear{{Sch{\"o}nrich} \& {Binney}}{{Sch{\"o}nrich}
  \& {Binney}}{2009a}]{SB09a}
{Sch{\"o}nrich} R.,  {Binney} J.,  2009a, MNRAS, 396, 203

\bibitem[\protect\citeauthoryear{{Sch{\"o}nrich} \& {Binney}}{{Sch{\"o}nrich}
  \& {Binney}}{2009b}]{SB09b}
{Sch{\"o}nrich} R.,  {Binney} J.,  2009b, MNRAS, 399, 1145

\bibitem[\protect\citeauthoryear{{Seigar} \& {James}}{{Seigar} \&
  {James}}{1998}]{SJ98}
{Seigar} M.~S.,  {James} P.~A.,  1998, \mnras, 299, 685

\bibitem[\protect\citeauthoryear{{Sellwood}}{{Sellwood}}{2014}]{Sellwood13}
{Sellwood} J.~A.,  2014, Reviews of Modern Physics, 86, 1

\bibitem[\protect\citeauthoryear{{Sellwood} \& {Binney}}{{Sellwood} \&
  {Binney}}{2002}]{SB02}
{Sellwood} J.~A.,  {Binney} J.~J.,  2002, MNRAS, 336, 785

\bibitem[\protect\citeauthoryear{{Sellwood} \& {Carlberg}}{{Sellwood} \&
  {Carlberg}}{2014}]{SC14}
{Sellwood} J.~A.,  {Carlberg} R.~G.,  2014, \apj, 785, 137

\bibitem[\protect\citeauthoryear{{Solway}, {Sellwood} \&
  {Sch{\"o}nrich}}{{Solway} et~al.}{2012}]{SSS12}
{Solway} M.,  {Sellwood} J.~A.,    {Sch{\"o}nrich} R.,  2012, \mnras, 422, 1363

\bibitem[\protect\citeauthoryear{{Spitzer} Jr. \& {Schwarzschild}}{{Spitzer} \&
  {Schwarzschild}}{1953}]{SS53}
{Spitzer} Jr. L.,  {Schwarzschild} M.,  1953, \apj, 118, 106

\bibitem[\protect\citeauthoryear{{Vera-Ciro}, {D'Onghia}, {Navarro} \&
  {Abadi}}{{Vera-Ciro} et~al.}{2014}]{Vera-Ciro14}
{Vera-Ciro} C.,  {D'Onghia} E.,  {Navarro} J.,    {Abadi} M.,  2014, \apj, 794,
  173

\bibitem[\protect\citeauthoryear{{Wielen}}{{Wielen}}{1977}]{Wielen77}
{Wielen} R.,  1977, \aap, 60, 263

\end{thebibliography}

\appendix
\section{Two Capture Criteria}\label{sec:AppxA}

In the literature, there are two separate criteria \citep{BT87,Contopoulos78} for whether or not a star with zero random orbital energy ($E_{ran}=0$) is in a trapped orbit.  The maximum radial excursion for a trapped orbit used by \cite{SB02} (their eqn.~12) comes from the the equations of motion used to derive the capture criterion in \cite{BT87}.  This is not the criterion we discuss in \S\ref{sec:CaptureWithout} of this work.  In this appendix, we outline the set of assumptions used in the derivation of each capture criterion.  

\subsection{Criterion used by \citeauthor{SB02}}\label{sec:AppxA1}

The capture criterion in \cite{BT87} (Chapter 3.3b) is derived using perturbation theory.  The disk potential ($\Phi(R,\phi)$) is assumed to be composed of an underlying axi-symmetric potential ($\Phi_0(R)$) plus an $m$-armed perturbation to the potential ($\Phi_1(R,\phi)$).  The effective potential, $\Phi_{eff}$  (eqn.~\ref{eqn:EffectivePotential}), is the potential in a frame that rotates with the pattern speed of the perturbation to the potential ($\Omega_p$).  Following the derivation from \cite{BT87}, a minimum in the effective potential at the radius of corotation ($R_{CR}$, where $\Omega(R)=\Omega_p$) is located at azimuth given by $\phi=0$.  The local maxima in the effective potential (between the arms of the perturbation) are at the radius of corotation and azimuth given by $\phi=\pi/m$ and at consecutive intervals every $2\pi/m$. 

It is assumed that the guiding centre of a star in a trapped orbit is located at a local maximum in the effective potential at position $(R_0$,$\phi_0)=(R_{CR}$,~$\pi/m)$.  It is further assumed that the equations of motion for a star in a trapped orbit can be described as small, oscillatory excursions around its guiding centre such that $R_1(t)$ is the time dependent radial distance from the radius of corotation ($R_1(t)=R(t)-R_{CR}$), and the azimuthal excursions are described by $\psi(t)=m(\phi(t)-\phi_0)$.  Orbital motions around this guiding centre are thus assumed to be similar to an epicyclic excursion in that there is no change in angular momentum.

They show that there is an integral of motion in the rotating frame for a star in a trapped orbit that is given by \citep[][eqn.~3.128]{BT87},
\begin{equation}\label{eqn:A1Ep}
E_p = \frac{1}{2}\dot{\psi}(t)^2-p^2 \cos\psi(t) ,
\end{equation}
where \citep[][eqn.~3.127b]{BT87}
\begin{equation}
p^2 = |\Phi_b(R_{CR})|\dfrac{4}{R_{CR}^2}\dfrac{(4\Omega_{CR}^2-\kappa_{CR}^2)}{\kappa_{CR}^2}
\end{equation}
and $|\Phi_b(R_{CR})|$ is the amplitude of the perturbation to the potential evaluated at the radius of corotation and $\Omega_{CR}$ and $\kappa_{CR}$ are the circular and orbital frequencies of a star evaluated at $R_0=R_{CR}$ in the underlying axi-symmetric potential.  In eqn.~\ref{eqn:A1Ep}, $\frac{1}{2}\dot{\psi}(t)^2$ is the $\hat{\phi}$-directional kinetic energy in the rotating frame and $-p^2\cos\psi(t)$ is the potential in this regime.  A star with $E_p<p^2$ will oscillate with simple harmonic motion in the potential described by $-p^2\cos\psi$, whereas a star that does not satisfy this criterion will circulate around the galactic centre in the rotating frame.

We re-express $E_p<p^2$ by using eqn.~\ref{eqn:A1Ep}, so that,
\begin{equation}
\frac{1}{2}\dfrac{\dot{\psi}(t)^2}{(1+\cos\psi(t))} < p^2 .
\end{equation}

By rearranging some constants and substituting for $p$,
\begin{equation}\label{eqn:A1midstep}
\frac{1}{2}(4\Omega_{CR}^2-\kappa_{CR}^2)\dfrac{\kappa_{CR}^2 }{4 \Omega_{CR}^2}\dfrac{R_1(t)^2}{(1+\cos\psi(t))} < |\Phi_b(R_{CR})| .
\end{equation}

In a power law potential (e.g.,~eqn.~\ref{eqn:BetaPotentials}), we can write the relation $\gamma^2\Omega^2=\kappa^2$, where the value of gamma must satisfy $1\leq\gamma\leq 2$ ($\gamma$ is related to $\beta$ from \S\ref{sec:CaptureWithEpi} by $\gamma=\sqrt{4-\beta}$).  Under this assumption, eqn.~\ref{eqn:A1midstep} can be expressed, 
\begin{equation}\label{eqn:BTcapture}
\frac{1}{2}\dfrac{(4-\gamma^2)\gamma^2}{4}\Omega_{CR}^2\dfrac{(R(t)-R_{CR})^2}{(1+\cos\psi(t))} < |\Phi_b(R_{CR})| ,
\end{equation}
where we have replaced $R_1(t)$ with $R(t)-R_{CR}$ as defined at the beginning of this subsection.  In a disk with a flat rotation curve ($\gamma^2=2$), a star in libration around the local maximum in the effective potential must satisfy
\begin{equation}\label{eqn:A1Capture}
\frac{1}{2}\Omega_{CR}^2\dfrac{(R(t)-R_{CR})^2}{(1+\cos\psi(t))} < |\Phi_b(R_{CR})| .
\end{equation}

A star that has a large value for $R(t)-R_{CR}$ may not satisfy eqn.~\ref{eqn:BTcapture} and will circulate about the galactic centre with a guiding centre radius $R_0 \neq R_{CR}$.  Indeed, eqn.~\ref{eqn:BTcapture} cannot be used to evaluate whether or not a star is in a trapped orbit if $\Omega_0 \neq \Omega_{CR}$ because of the initial assumption that $R_0=R_{CR}$ and that all non-circular motions come from oscillations about the star's guiding centre, which is located at the local maximum in the effective potential. 

The capture criterion derived in \cite{BT87} can be used only for stars with zero random energy.  It predicts that a star will either meet the capture criterion or be on a circular orbit about the galactic centre.  Non-circular motions are produced by a star that meets the capture criterion as it oscillates about its guiding centre.

\subsection{Criterion derived by \citeauthor{Contopoulos78}}\label{sec:AppxA2}

The capture criterion discussed in \S\ref{sec:CaptureWithout} and derived by \cite{Contopoulos78} assumes that the star being evaluated has zero radial action, corresponding to zero random orbital energy by eqn.~\ref{eqn:SB02JR}.  In this method, the motions of a star in a trapped orbit around the local maximum in the effective potential are caused by changes in that star's angular momentum.  Therefore, $R(t)=R_L(t)$.  \citeauthor{Contopoulos78}'s capture criterion (eqn.~\ref{eqn:trapped}) has both an upper and lower limit that must be satisfied in order for the star to be in a trapped orbit and can be applied to all disk stars that have zero random orbital energy.  The lower limit,
\begin{equation}\label{eqn:A2Capture}
\Lambda_c>-1
\end{equation} 
determines the boundary between stars in trapped orbits and stars that circulate about the galactic centre in the rotating frame.  We use this limit as the statement that describes a scenario similar to the capture criterion in \S\ref{sec:AppxA1} ($E_p<p^2$).  One should keep in mind that the capture criterion we now discuss (eqn.~\ref{eqn:A2Capture}) states whether or not a star with any given angular momentum (and therefore any given guiding centre radius) is in a trapped orbit; whereas, it is assumed in \S\ref{sec:AppxA1} that the star has a guiding centre radius at $R_{CR}$.  

We can evaluate eqn.~\ref{eqn:A2Capture} by using the same perturbation to the potential as in \S\ref{sec:AppxA1}.  The value of $\Lambda_c$ is given by eqn.~\ref{eqn:Lambda_c}.  Its evaluation requires one to calculate the Jacobi integral of the star, $E_J$, given by eqn.~\ref{eqn:EJ}.  The velocity of a star in the rotating frame is given by $\mathbf{{x}_\phi}(t) = R_L(t)(\Omega_c(R_L(t))-\Omega_{CR})\boldsymbol{\hat{\phi}}+\mathbf{v}_{ran}$.  In the limit that $\mathbf{v}_{ran}\rightarrow 0$ (zero random energy), we find that the capture criterion can be expressed as,
\begin{equation}\label{eqn:A2CaptureLong}
\dfrac{\left[ \Phi_0(R_L(t))-\Phi_0(R_{CR}) \right] + \frac{1}{2}R_L(t)^2 |\Omega_c(R_L(t))-\Omega_{CR}|^2 +\frac{1}{2}\Omega_{CR}^2( R_{CR}^2-R_L(t)^2)}{(1+\cos\psi(t))} < |\Phi_b(R_{CR})| .
\end{equation}

In a flat rotation curve and in the limit that $\Phi_0(R(t))\approx\Phi_0(R_{CR})$ and $\Phi_1(R_L(t))\approx\Phi_1(R_{CR})$, we find,
\begin{equation}\label{eqn:A2CaptureReduced}
\Omega_{CR}^2 \dfrac{R_{CR}|R_L(t)-R_{CR}|}{(1+\cos \psi(t))} < |\Phi_b(R_{CR})| .
\end{equation}

The criterion in eqn.~\ref{eqn:A2CaptureReduced} is neither an exact mathematical nor physical equivalent to the criterion in eqn.~\ref{eqn:A1Capture}.  In the current case (eqn.~\ref{eqn:A2CaptureReduced}) the star has azimuthal motion that is associated with a circular orbit around the galactic centre, where $R(t)=R_L(t)$ and $\Omega_c(R_L(t))=v_c/R_L(t)$.  This method does not assume a priori that $R_L=R_{CR}$, as is the case in \S\ref{sec:AppxA1}.

\label{lastpage}

\end{document}